\documentclass[10pt,a4paper]{article}
\usepackage[latin1]{inputenc}
\usepackage[english]{babel}
\usepackage{amsmath}
\usepackage{amsfonts}
\usepackage{amssymb}
\usepackage{graphicx}
\usepackage{authblk}
\usepackage{ulem}
\usepackage[title]{appendix}

\title{Modelling time-varying interactions in complex systems: the Score Driven Kinetic Ising Model}
\author[a,c,d]{Carlo Campajola
\footnote{Corresponding author: carlo.campajola@uzh.ch}}
\author[a]{Domenico Di Gangi} 
\author[a,b]{Fabrizio Lillo}
\author[b]{Daniele Tantari}

\affil[a]{Scuola Normale Superiore, p. dei Cavalieri 7, 56126 Pisa, Italy}
\affil[b]{Department of Mathematics, University of Bologna, p. di Porta San Donato 5, 40126 Bologna, Italy}
\affil[c]{UZH Blockchain Center, University of Z\"{u}rich, R\"{a}mistrasse  71,  8006  Z\"{u}rich, Switzerland}
\affil[d]{Institute of Informatics, University of Z\"{u}rich, Andreasstrasse 15, 8050 Z\"{u}rich, Switzerland}

\begin{document}

\maketitle

\begin{abstract}
    A common issue when analyzing real-world complex systems is that the interactions between the elements often change over time: this makes it difficult to find optimal models that describe this evolution and that can be estimated from data, particularly when the driving mechanisms are not known. Here we offer a new perspective on the development of models for time-varying interactions introducing a generalization of the well-known Kinetic Ising Model (KIM), a minimalistic pairwise constant interactions model which has found applications in multiple scientific disciplines. Keeping arbitrary choices of dynamics to a minimum and seeking information theoretical optimality, the Score-Driven methodology lets us significantly increase the knowledge that can be extracted from data using the simple KIM. In particular, we first identify a parameter whose value at a given time can be directly associated with the local predictability of the dynamics. Then we introduce a method to  dynamically learn the value of such parameter from the data, without the need of specifying parametrically its dynamics. Finally, we extend our framework to disentangle different sources (e.g. endogenous vs exogenous) of predictability in real time. 
We apply our methodology to several complex systems 
including financial markets, temporal (social) networks, and neuronal populations. Our results show that the Score-Driven KIM produces insightful descriptions of the systems, allowing to predict forecasting accuracy in real time as well as to separate different components of the dynamics. This provides a significant methodological improvement for data analysis in a wide range of disciplines.
\end{abstract}

\section{Introduction}

Complex systems, characterized by a large number of simple components that interact with each other in a non-linear way, have been an increasingly important field of study over the last decades. Interactions make the whole more than the sum of its parts \cite{bar2002general}: for this reason the effort when modeling complex systems is ultimately directed to understand how interactions arise, how to parametrize them into quantitative models and how to estimate them from empirical measurements. 

One complication that is ubiquitous to real complex systems, but very rarely considered in modeling, is that interactions change over time: traders in financial markets continuously adapt their strategic decision-making to each other's actions \cite{challet2016trader} and to new information \cite{lillo2015news}; preys change their behavior to avoid predators \cite{schmitz2017predator}; neurons reinforce (or inhibit) connections in response to stimuli \cite{tavoni2017functional}. As we show also below, a modeling approach assuming that all the interactions are constant can be misleading, sometimes leading to spurious estimations of the interactions, which can be avoided only with very strong limitations to sample selection and experimental design (when possible).

In this article we propose a novel approach to the development of models for time-varying interactions based on the generalization of a minimalistic constant-interactions model, which is commonly used in many scientific disciplines, the Kinetic Ising Model (KIM) \cite{crisanti1988dynamics}. In the following we show that this generalization allows to describe conditions where the predictability of the overall dynamics of the observed process is variable, while commonly employed constant interaction models fail in this respect. More importantly, our modeling approach does not assume that the causes or the dynamics of the variable interactions are known, but they are estimated (or filtered) from the data themselves. Thus, different types of time-varying interactions can be present in the investigated system, including non-stationarities of various form (regime-shift, seasonalities, etc.).
Indeed it often occurs that the modeler has no insight on the nature of the underlying dynamics of interactions: the dynamics that is given to the time-varying parameters then needs to be as agnostic as possible with respect to the actual generating dynamics, {\it i.e.} be robust to model misspecification errors.

Since it is generally difficult  to determine why and how interactions change over time,  it is even riskier to try to model their dynamics with specific external drivers. Conversely we assume a generic Markovian paradigm in which $J_{ij}(t)$ - representing the interaction between the system's  variables $s_i(t)$ and $s_j(t)$ at time $t$ - endogenously adapt to the observations of  $s(t)$ themselves, i.e.  
\begin{equation}
    J_{ij} (t+1) = F (J(t),s(t)).
\end{equation}
The updating functional $F$ is determined by general assumptions based on information theory principles. First of all, one can assume the interactions variation depends on surprise: the more an observation of the system's state is ``unexpected'',  the more the relations between its components will change.  In social systems,  for example, friendship relationships can get damaged if not constantly fed or may arise from unexpected  gestures of openness. This is also a common principle in biological learning processes and artificial neural networks, where the least expected inputs have the largest impact on the values of the synapses or inter-units weights \cite{ackley1985learning}.  The most widespread measure of surprise is minus the logarithm of the conditional likelihood $p(s(t)| J(t))$ of observing a given pattern with the current level of interactions.  As a second principle we assume that the system's reaction to surprise is to adapt to it, making what has been unexpected for that moment,  expected for the future. In this sense the interactions change to increase the log likelihood of the last observation i.e. 
\begin{equation}
    J(t+1) = w + B J(t) + A(t) \frac{\partial \log p (s(t) \vert J(t))}{\partial J(t)},
\end{equation}
which can be interpreted as the updating rule of an autoregressive process with a gradient ascent perturbation  with given learning rate parameter $A(t)$, which possibly depends on time.
This type of  observation-driven \cite{cox1981statistical} dynamics has been recently introduced \cite{creal2013generalized, harvey_2013} in defining the class of {\it score-driven models}. These have been shown to be an optimal choice among observation-driven models when minimizing the Kullback-Leibler divergence to an unknown generating probability distribution \cite{blasques2015information} and have risen in popularity in econometrics \cite{bernardi2019switching} as well as network science \cite{di2019score}.

In general, given a sequence of observations $\lbrace s(t) \rbrace_{t=1}^T$ where $s(t) \in\mathbb{R}^N$ and a model with conditional log-likelihood $\mathcal{L}(t) = \log p (s(t) \vert \mathcal{S}(t-1), f(t) )$ depending on a vector of time-varying parameters $f(t) \in \mathbb{R}^M$ and past observations $\mathcal{S}(t-1) = \lbrace s(k) \rbrace_{k = 1}^{t-1}$,
a score-driven model assumes that the time evolution of $f(t)$ is ruled by the recursive relation

\begin{equation}  \label{eq:gasupdaterule}
f(t+1) = w + B f(t) + A \mathcal{I}^{-1/2}(t) \frac{\partial \mathcal{L}(t)}{\partial f(t)}
\end{equation}
where $w$, $B$ and $A$ are a set of static parameters. In the rest of the article we will call $\nabla_t = \frac{\partial \mathcal{L}(t)}{\partial f(t)}$ the \textit{score function} at time $t$, hence the name \textit{score-driven model}. $\mathcal{I}^{-1/2}(t)$ is a $M\times M$ matrix regularizing the convexity, that we choose to be the inverse of the square root of the Fisher information associated with $\mathcal{L}(t)$, thus letting the last term of Eq. \ref{eq:gasupdaterule} be a random variable with unit variance and zero mean by definition.

As is clear from Eq. \ref{eq:gasupdaterule}, the score $\nabla_t$ drives the time evolution of $f(t)$ and no additional source of noise is introduced. This means that, given a $p(s(t) \vert \mathcal{S}(t-1), f(t))$, the sampling of the observations from this distribution produces a deterministic update of the time-varying parameters. The update can remind the reader of a Newton-like method for optimization, in that the parameters are moved towards the maximum of the likelihood at each realization of the observations while keeping memory of the time evolution through the $B$ static parameter.

The fact that  time-varying parameters are deterministic functions of the observations has some intrinsic advantages also for estimation, as the elimination of unobservable noise removes the necessity of implementing computationally intensive Monte Carlo simulations to calculate the model likelihood. Furthermore, an observation-driven model can be used as a filter: having knowledge of all the static parameters ({\it e.g.} because they were previously estimated on a training set), the time-varying parameters can be updated with no effort every time a new data point is observed. In the following we will make wide use of the score-driven model as filter of a unknown dynamics and in the SI we provide more intuition of it by revisiting the simple case of a GARCH process \cite{nelson}.

The focus of the paper is the score-driven generalization of the Kinetic Ising Model (KIM) \cite{derrida1987exactly,crisanti1988dynamics}, which is the dynamical counterpart of the celebrated Ising spin glass model \cite{kirkpatrick1978infinite, edwards1975theory}. 
Ising models in general are known to be among the simplest models of complex systems that have been developed in the field of statistical physics and are at the roots of the theory on collective behavior and phase transitions. This large interest is also due to the fact that they fall into the class of Maximum Entropy models \cite{jaynes1957information, schneidman2006weak, marre2009prediction} when only average values and cross correlations are taken into account. The KIM in particular has been adopted in a variety of fields, such as neuroscience \cite{cocco2017functional, nghiem2018maximum, ferrari2018separating}, computational biology \cite{tanaka1977model,imparato2007ising,agliari2011thermodynamic}, economics and finance \cite{bornholdt2001expectation, bouchaud2013, sornette2014physics, campajola2020unveiling} and has been studied in the literature of machine learning \cite{lecun2015deep,hornik1989multilayer, decelle2015inference} to understand recurrent neural network models.

The KIM describes the time evolution of a set of $N$ binary variables $s(t) \in \lbrace -1, 1 \rbrace^N$ for $t = 1, \dots, T$, typically called ``spins'', which can influence each other through a time lagged interaction. We focus on its applications to time series analysis and extend it to allow the presence of time-varying parameters with score-driven dynamics.

In its standard form the Kinetic Ising Model for time series \cite{campajola2019inference} involves three main sets of parameters: a $N \times N$ interaction or coupling matrix $J$ and a $N$-dimensional vector $h$ of variable-specific biases, which we summarize as $\Theta = (J,h)$. The model is Markovian with synchronous dynamics, characterized by the transition probability

\begin{eqnarray}
p(s(t) \vert s(t-1); \beta, \Theta) = \frac{e^{\beta \sum_i s_i(t) g_i(t)} }{Z(t)}
\label{eq:transprob}
\end{eqnarray}
where $Z(t)$ is a normalizing constant commonly known as the partition function in statistical mechanics, and $\beta$ is a parameter that determines the amount of noise in the dynamics, known as the \textit{inverse temperature}. Typically the quantity $
g_i(t)\equiv  \sum_j J_{ij} s_j(t-1) + h_i$ is called the \textit{effective field} perceived by spin $i$ at time $t$. Furthermore, it is possible in principle to introduce dependency on any number $K$ of external regressors $x_k(t)$, by adding a term $b_{ik}x_k(t)$ to $g_i(t)$ for each $k \in \lbrace 1, \dots, K \rbrace$, as done for instance in \cite{campajola2020unveiling}. From the standard KIM we use Eq.(\ref{eq:gasupdaterule}) to provide a dynamics to the parameters $(\beta,\Theta)$ thus introducing a Score Driven generalization of the  KIM. 
Notice however that the number of parameters in the KIM is large, $O(N^2)$: as customary in high-dimensional modeling, in the following we will propose two parsimonious and informed parameter restrictions that simplify the treatment and define two kinds of Score-Driven KIM, each tailored to highlight different effects.

As we show in this article, the development of a score-driven KIM addresses three important points: first, introducing a dynamical noise parameter $\beta(t)$ allows to gain real time insight on the ability of the model to explain the observed dynamics, thus leading to more informed forecasts; second, neglecting time variability of parameters by estimating a standard KIM turn out to produce systematic errors, in particular the estimated values are different from the time-averaged values that generated the sample;  third, by introducing a convenient factorization for the model parameters, it is possible to discriminate whether an observation is better explained by endogenous interactions with other variables or by exogenous effects, offering an improved understanding of the dynamics that generated the data even when these effects are not constant over time. We prove the effectiveness of our modeling approach by extensive numerical simulations and by empirical application to different complex systems.

\section{Results}

\subsection{The Dynamical Noise KIM}\label{sec:dyno}

The first score-driven KIM we propose addresses the first two points made above, namely the real time prediction of forecast accuracy and the correction of systematic estimation errors of a constant parameter model. The Dynamical Noise KIM (DyNoKIM) is defined by letting the noise parameter $\beta$ in Eq. \ref{eq:transprob} be time-varying, while all other parameters are constant. 

\begin{figure}[t]
    \centering
    \includegraphics[width=.9\linewidth]{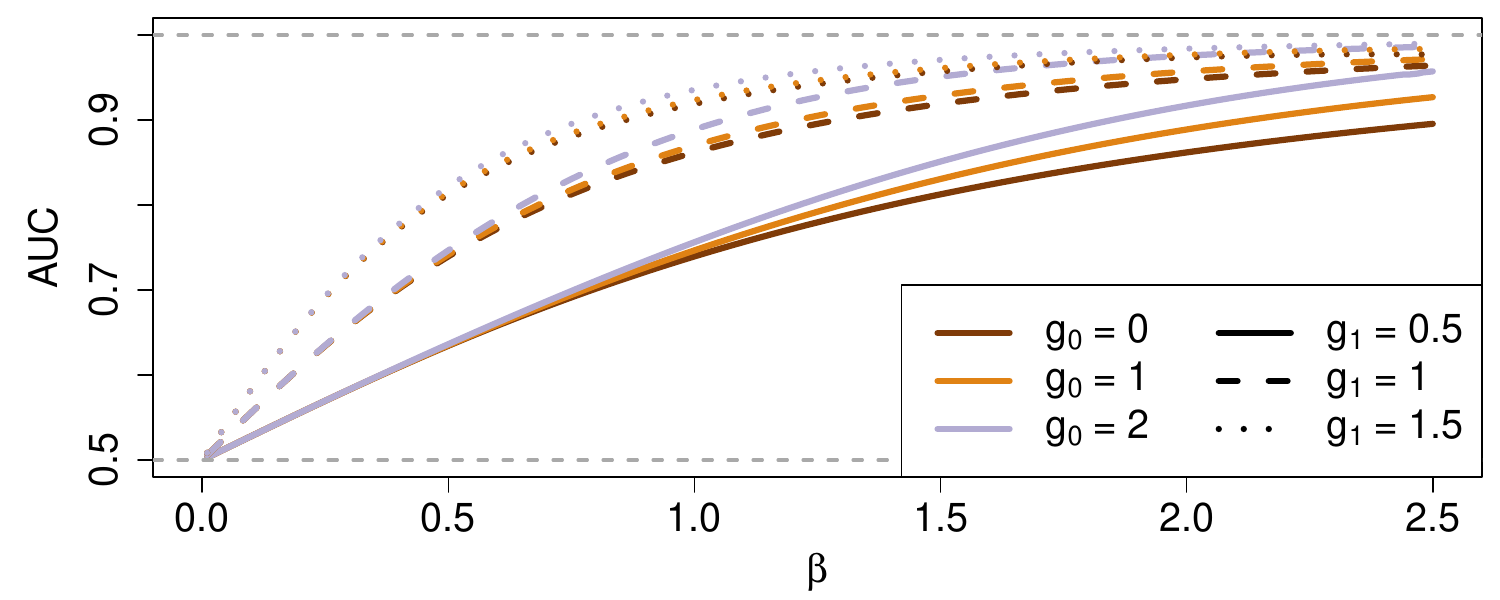}
    \caption{Theoretical AUC as a function of $\beta$ assuming $g_i$ is Gaussian distributed with mean $g_0$ and standard deviation $g_1$. Different colors correspond to different values of $g_0$, while line types identify values of $g_1$. We see that increasing $\beta$ has the effect of reducing the uncertainty on the random variable $s_i(t)$, keeping $g_i$ unchanged. Grey dashed lines at $\mathrm{AUC}=0.5$ and $\mathrm{AUC}=1$ are guides to the eye.}
    \label{fig:entropy}
\end{figure}

To better understand the rationale behind this choice, let us introduce the theoretical Area Under the ROC Curve (AUC) \cite{hanley1982meaning, bradley1997use}, a standard measure of the accuracy of the forecast, and study how it varies as a function of $\beta$ in the standard KIM. We provide the details of the derivation in the SI where we show how the AUC depends both on $\beta$ and on the unconditional distribution of the effective fields $g_i(t)$.
In Figure \ref{fig:entropy} we display the result assuming that $g$ is Gaussian distributed with mean $g_0$ and standard deviation $g_1$. This is the case for instance if the $J_{ij}$ entries are Gaussian distributed with zero mean. We see that the AUC is monotonically increasing with $\beta$, but also that the distribution of the static parameters affects the slope with which the curve converges towards 1, namely the smaller the mean and variance of the effective fields $g_i$, the slower the growth of AUC. Figure \ref{fig:entropy} tells us that the larger is $\beta$ the more reliable is the prediction of the model. Hence if we are able to estimate $\beta$ locally we can assess in real time how good the model is in forecasting the next observation. This is why in the DyNoKIM we consider a time-varying $\beta$.

Specifically, the DyNoKIM is characterized by the transition probability 

\begin{equation}
p(s(t) \vert s(t-1); J, \beta(t))
= \frac{ e^{ \beta(t) \sum_{i} s_i(t) g_i(t)}}{Z(t) }
\label{eq:dynokim}
\end{equation}

\noindent with $Z(t) = \prod_i 2 \cosh \left[\beta(t) g_i(t)\right]$. We give score-driven dynamics to $f(t) = \log \beta(t)$, as $\beta$ is positive and inversely related to the noise:

\begin{equation}
\log\beta(t+1) = w + B \log\beta(t) + A \mathcal{I}^{-1/2}(t) \nabla_t
\label{eq:betadynamics}
\end{equation}
where $w$, $B$ and $A$ are scalar parameters and $\mathcal{I}(t)$ is the Fisher Information and $\nabla_t = \frac{\partial p(s(t) \vert s(t-1), \beta(t))}{\partial \beta(t)}$ is the score. 

The interpretation for this model is simple yet extremely useful: the higher the value of $\beta$, the smaller the uncertainty over the realization of $s(t)$ or, in other words, the more accurate a prediction of the value of $s(t)$, as we have shown in Fig. \ref{fig:entropy}. Operationally, at a given time $t-1$ with an observation $s(t-1)$, it is possible to use the DyNoKIM to produce one-step ahead forecasts for $s(t)$, which we call $\hat{s}_i(t)$. These are obtained as 

\begin{equation}\label{eq:forecast}
    \hat{s}_i(t) = \mathrm{sign} \left[p \left( s_i(t) = 1 \big\vert s(t-1), \Theta, \beta(t) \right) - \alpha \right]
\end{equation}
where $\alpha$ is an arbitrary threshold level. Sweeping the value of $\alpha$ between 0 and 1 one obtains a ROC curve, which in turn can be used to calculate the AUC. We report simulation results for this procedure in the SI. Notice that $\beta(t)$ depends only from past observations $\mathcal{S}(t-1)$ through Eq.\ref{eq:betadynamics}, thus the predictions are fully causal.

In the statistical physics literature there have been few attempts to study similar models \cite{penney1993coupled, beck2003superstatistics, beck2005time}. However these works assume that the sampling of the observations and of the time-varying parameters take place on two separated time scales, meaning that the parameters are locally constant when the observations are sampled. This is not true for score-driven models, which are in fact designed to not require this assumption, intuitively formalized by the values of the parameters $B$ and $A$. If $B \gg A$ then the evolution of $f$ is indeed slower than the one of observations, while if $B \ll A$ they evolve on the same time scale.

\begin{figure}[t]
    \centering
    \includegraphics[width=.9\linewidth]{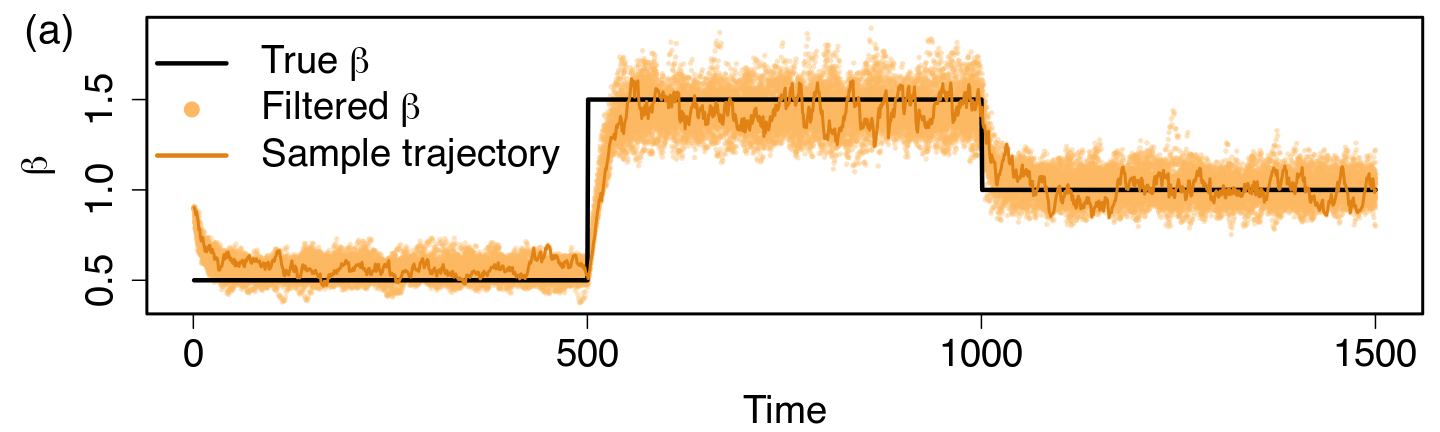}\\
    \includegraphics[width=.9\linewidth]{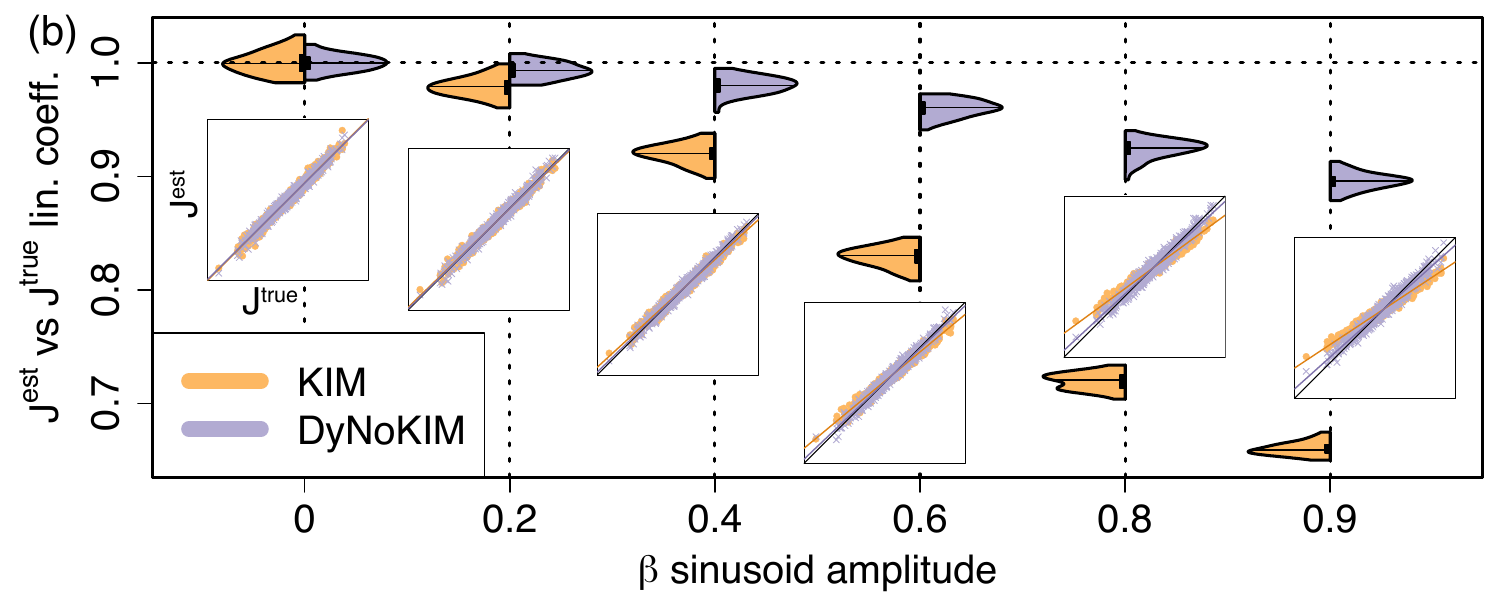}
    \caption{(a) Example of a filtered $\beta$ trajectory with a piecewise constant generating process over 30 simulations. (b) Estimation of $J$ under model misspecification with a time-varying $\beta(t) = 1 + \mathcal{K} \sin (\omega t)$, comparing the KIM and the DyNoKIM. On the $x$ axis we plot the amplitude $\mathcal{K}$, on the $y$ axis the distribution of the coefficient $b$ of the linear regression over 60 simulations. Insets show example scatter plots of the true $J$ values ($x$ axis) and the estimated values ($y$ axis) using the standard KIM (yellow points) or the DyNoKIM (purple crosses). Simulation parameters are $\omega = 2\pi/300$, $T=3000$, $N=30$, $J_{ij} \sim \mathcal{N}(0, 1/\sqrt{N})$, $h_i = 0 \; \forall \, i$ }
    \label{fig:misspec_doublestep}
\end{figure}

The estimation of the DyNoKIM requires some restrictions. It is known \cite{mezard2011exact} that, given a set of observations, the parameter $\beta$ in the standard KIM of Eq. \ref{eq:transprob} is not identifiable. In fact, for any two values $\beta_1$ and $\beta_2$ there are also two sets of parameters $\Theta_1$ and $\Theta_2$ such that $p(s(t) \vert s(t-1);\beta_1, \Theta_1) = p(s(t) \vert s(t-1); \beta_2, \Theta_2)$ for all $s(t)$. For this reason in inference problems it is typically assumed that $\beta=1$ incorporating its effect in the size of the other parameters. When $\beta$ is made time-varying though, the identification problem is limited to its time average value $\langle \beta \rangle$ (which still needs to be assumed equal to 1), while its local value can be inferred from the data. This result has implications particularly for forecasting applications: a forecast should be considered more or less reliable by looking at the value of $\beta(t)$ at the previous instant in time and considering how much above $0.5$ the corresponding expected AUC is, according to the relation shown in Figure \ref{fig:entropy}. Finally, the parameters of Eq. \ref{eq:betadynamics} are inferred by Maximum Likelihood Estimation (see Material and Methods). We numerically find that the model parameters can be consistently estimated and report a detailed analysis in the SI. 

Our main focus here is to study the model's ability to retrieve the correct parameters also when the data generating process is not score-driven. Indeed there is little reason to believe that this sort of dynamics is an actual data generating process for real-world complex systems, where $\beta$ might follow exogenous and unknown dynamics.  The power of score-driven models lies also in the capability of estimating time-varying parameters, such as $\beta(t)$, without actually requiring any assumption on their true dynamical laws. In this sense they behave as filters for the underlying unknown dynamics of the parameters. To show that this is the case also for the DyNoKIM, Fig. \ref{fig:misspec_doublestep}a displays an example of misspecified $\beta(t)$ dynamics, a deterministic double step function, that is correctly recovered by the score-driven approach. We simulate 30 time series of length $T$ using the given values of $\beta(t)$ to generate the $s(t)$; given only the simulated $s(t)$ time series, the inference algorithm determines the optimal static parameters $A$, $B$ and $J$ and filters the optimal value of $\beta(t)$ at each time. The resulting $\beta^{est}(t)$ values are well localized around the simulated ones.

One could argue that a KIM with a time varying $\beta(t)$ has similar performances to a standard KIM with a constant $\beta$ equal to $\langle \beta \rangle$. This is not the case.  
Fig. \ref{fig:misspec_doublestep}b shows the results for a set of simulations where $\beta(t)$ follows a deterministic sinusoidal dynamics, $\beta(t) = 1 + \mathcal{K} \sin \omega t$, varying the amplitude $\mathcal{K}$, and the time evolution of $s(t)$ is given by Eq. \ref{eq:dynokim}. For each value of $\mathcal{K}$ we simulate 60 time series of $T$ observations and fit both the constant parameters KIM and the score-driven DyNoKIM, then comparing the inferred $J^{est}$ with the one that was used to generate the data, $J^{true}$, by means of a linear regression model $J^{est}_{ij} = a + bJ^{true}_{ij} + \epsilon$. We see from Figure \ref{fig:misspec_doublestep}b that when $\beta$ is not constant, the KIM underestimates the absolute value of the parameters, highlighted by the fact that $b < 1$ (and $a \approx 0$, not shown). The error is greatly reduced in the DyNoKIM thanks to the way in which we solve the indetermination of $\langle \beta \rangle$: after the model parameters are estimated and a filtered $\beta^{est}(t)$ is found, we normalize its mean to $1$ and multiply the estimated $J^{est}$ by the same factor, leaving the likelihood of the model unchanged. This result supports our argument that using a KIM on data where parameters of the data generating process are time varying can be misleading and leads to significant errors, something that can be overcome by adopting the score driven models proposed here. 

\subsection{Forecasting stock price activity with DyNoKIM}
The first application of DyNoKIM is to financial markets.
Measuring high-frequency price volatility in financial markets is a non-trivial task that has been at the core of research in quantitative finance over the last two decades \cite{ait2011ultra}. Volatility is in fact a latent process which is hard to measure for reasons that range from price staleness to microstructural effects like price discretization and bid-ask bounce. Price activity, namely the binary time series marking the events of price changes, is a proxy for high-frequency price volatility that has been used recently to quantify the endogeneity in the price formation \cite{filimonov2012quantifying, hardiman2013critical, filimonov2015apparent, hardiman2014branching, wheatley2019endo,rambaldi2015modeling, rambaldi2018detection}. 

Here we propose the DyNoKIM as an effective tool to forecast stock price activity at high frequency. The advantages with respect to standard methods is twofold: first, we are able to model the dynamics of a large panel of assets, hence considering  volatility spillovers between them; second, the score driven approach allows us to measure the local predictability of price activity in real time. 
We study the $100$ largest capitalization stocks in the NASDAQ and NYSE over 11 trading days. Price activity is defined as a binary variable $s_i(t)$ for each stock $i$, taking value $+1$ if the stock price has changed in the interval $(t-1,t]$ and $-1$ otherwise, with time discretized at $5$ seconds. The choice of time scale is largely arbitrary: we choose 5 seconds to obtain a set of variables that have unconditional mean as close to 0 as possible to have a balanced dataset. We focus our attention on the lagged interdependencies among different stocks, by applying the DyNoKIM to the multivariate time series $s(t)$.

\begin{figure}[t]
    \centering
    \includegraphics[width=.9\linewidth]{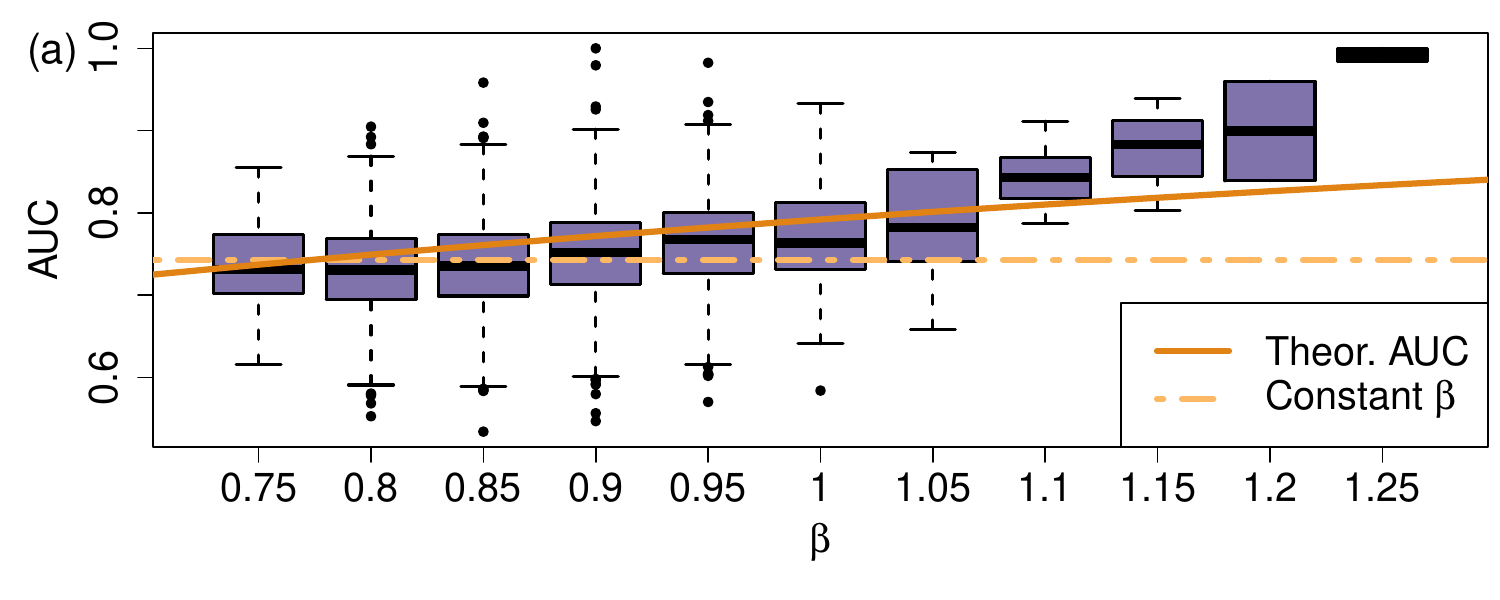}
    \includegraphics[width=.9\linewidth]{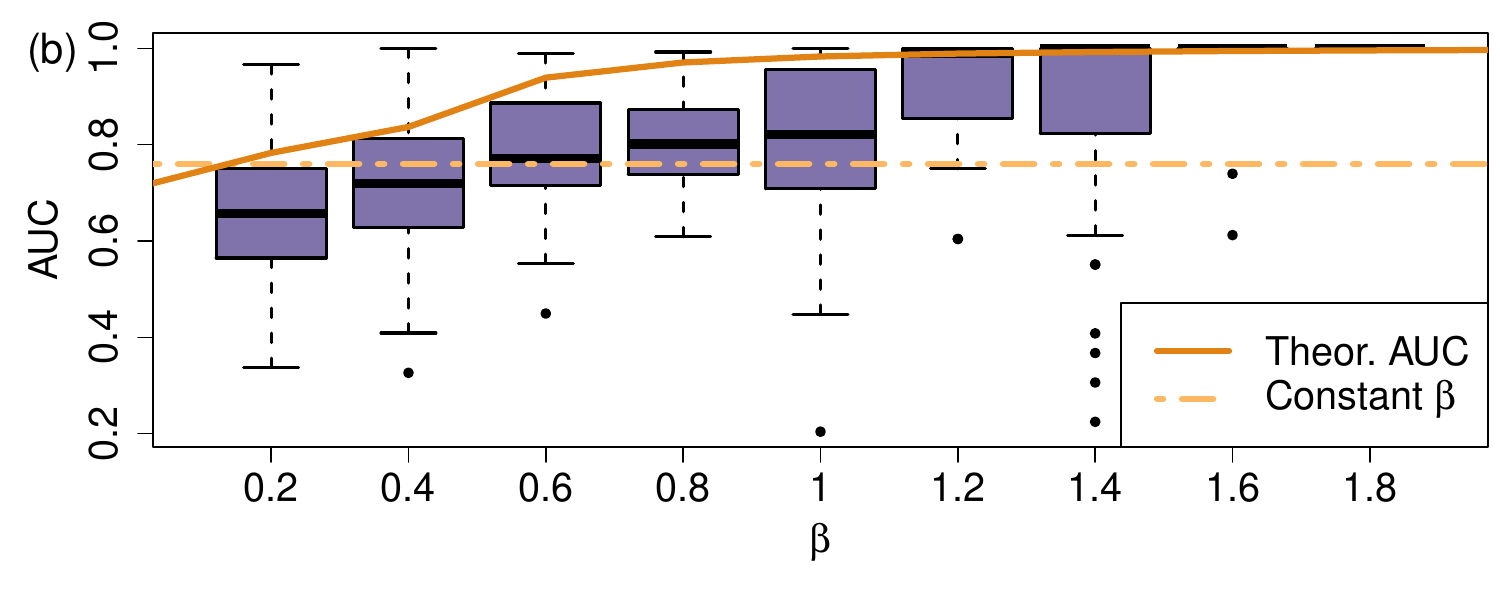}
    \caption{AUC statistics compared to $\beta(t)$ in applications of DyNoKIM. (a) AUC values for stock price activity on November 19, 2019 aggregated for different values of $\beta(t)$, compared to the theoretical expected AUC with Gaussian $g_i(t)$ and to the performance with constant $\beta$; 
    (b) AUC values for link prediction in the SocioPatterns dataset, compared with the theoretical expected AUC and the constant $\beta$ benchmark.}
    \label{fig:aucstats}
\end{figure}

Our theoretical results from Figure \ref{fig:entropy} suggest to use $\beta$ to quantify the reliability of forecasts of price activity using this model. We thus estimate the model parameters once per day and use them to filter $\beta(t)$ on the following day, while measuring the out of sample accuracy of the predicted price movements using the AUC metric. To ensure that there is reason to model the system with time-varying $\beta$, we apply a Lagrange Multiplier (LM) test \cite{calvori2017testing} with a null hypothesis of constant $\beta$, finding strong rejections of the null at the $p<0.001$ level for every day in the sample. Further information on the test can be found in the Materials and Methods section.

We show an example of the results of this analysis in Fig. \ref{fig:aucstats}a where we consider a single day. We empirically observe that when the filtered value of  $\beta(t)$ is large, the subsequent forecast of activity is systematically more reliable because AUC is larger. 
We find a good agreement between the empirical results and the theoretical values for AUC under the assumption of Gaussian effective fields $g_i$, even if some discrepancy is observable due to the non-Gaussianity of actual fields. 
Thus we conclude that the DyNoKIM can be effectively used to model high frequency volatility of a large portfolio of stocks and to measure in real time its level of predictability.

\subsection{Link Prediction in Temporal Networks with DyNoKIM}
 In our second application, we show that DyNoKIM can be used to model temporal networks. In particular we show that DyNoKIM dynamically provides the level of predictability of links of the network by exploiting again the relation between $\beta(t)$ and AUC.

Networks are a paradigmatic tool to describe pairwise relations in complex systems \cite{newman2006structure, cohen2010complex, barabasi2013network, newman2018networks} and applications include human mobility \cite{gao2013understanding}, migration \cite{fagiolo2013international}, disease spreading \cite{draief2010epidemics}, international trade \cite{bhattacharya2008international} and financial stability \cite{gai2011complexity, cimini2015systemic}, to mention a few. 
More recently, the increasing availability of time varying relational data stimulated a widespread and fast growing interest in the analysis of temporal networks \cite{holme2012temporal}. It also motivated the development of a number of models to describe the dynamics of temporal networks \cite{sewell2015latent, sewell2018simultaneous, mazzarisi2020dynamic,Hanneke_tergm_10.1214/09-EJS548}


A network, defined by a set of $M$ nodes and a set of links between pairs of nodes, can be described by an $M\times M$ binary adjacency matrix $G \in \lbrace 0, 1 \rbrace^{M\times M}$ 
, where $A_{ij} = 1 $ if a link between nodes $i$ and $j$ is present and $G_{ij} = 0 $ otherwise. When the relation described by the links is not directional, $G_{ij} = G_{ji}$ and the network is said to be undirected. We consider temporal networks where the number of nodes $M$ is fixed across multiple time steps and indicate the adjacency matrix of the graph at time $t$ by $G\left( t \right)$. 

In order to use the KIM to model a temporal network, we map the  elements of the adjacency matrix into spins, associating a present link to a spin $+1$ and an absent link to a spin $-1$. In this way we represent each adjacency matrix $G\left( t \right)$ as a vector $s(t) \in \lbrace -1, 1 \rbrace^N$ where $N = M(M-1)/2$, assuming the network to be undirected and without self loops. In light of this mapping, the matrix $J$ now captures the tendency of links to influence each other at lag one - for example the diagonal terms can be interpreted as measuring link persistence - while the elements of $h$ are associated with the idiosyncratic probability to observe a given link.

Interestingly, such a mapping highlights that (standard) KIM can be seen as belonging to the Temporal Exponential Random Graph Model (TERGM)\cite{Hanneke_tergm_10.1214/09-EJS548} family, as we discuss in the SI.  Moreover, it turns out that a large subset of possible TERGM specifications can be mapped into a KIM. Hence, the score driven KIM  that we propose here is an extension of the TERGM allowing its parameters to evolve in time. This frames DyNoKIM also as a contribution to the literature on network models with time varying parameters, alongside with a recent extension of a different, but related, family called Exponential Random Graphs \cite{holl1981} to its score driven version \cite{di2019score}.


The problem of link prediction in networks is very important and can be framed in different ways \cite{wang2015link, martinez2016survey}. For discrete time temporal networks, link prediction amounts to forecasting the presence of a link at time $t+1$ given the observations available up to time $t$. This is easily done with the KIM defining the forecast exactly as in Eq \ref{eq:forecast}.

We apply DyNoKIM to a real world temporal network describing close proximity between workers at the Institut National de Veille Sanitaire in Saint-Maurice \cite{genois2015data}. The data was  collected with the sensing platform developed by the SocioPatterns \cite{sociopatternswebsite} collaboration and describe situations of face-to-face proximity between pairs of workers lasting at least $20$ seconds. The observations cover $10$ working days, from June 24 to July 3, 2013. For each day, we construct the time series of adjacency matrices, at a frequency of $20$ seconds between 7:30 am and 5:30 pm. A link between two workers is present if they face each other at a distance less than $1.5$ meters and is absent otherwise.
As is often the case in real temporal networks, a large number of links is never, or very rarely, observed. Since for such trivial links the prediction problem is not interesting, and to keep the computational complexity to a reasonable level, we consider only the subset of the $100$ most active links in each day. For each day, we estimate the DyNoKIM on a training set consisting of the first $75\%$ of observations and then use the remaining $25\%$ for out of sample validation. For each $t$ we compute the AUC and report in Fig. \ref{fig:aucstats}b the aggregated  results for all days. As in the financial application, we observe a monotonically increasing  relation between $\beta(t)$ and AUC, indicating that DyNoKIM is a reliable tool to dynamically quantify forecast accuracy also in applications to temporal networks data. Also in this case, we observe a good agreement with the theoretical prediction, with differences explainable by the non Gaussianity of the estimated matrix $J$.

These two empirical examples show that our theoretical results for the DyNoKIM are indeed verified in realistic applications and that using this method - which we believe could be applied even to more sophisticated models - can result in a significant gain in the use of forecasting models, giving a simple criterion to discriminate when to trust (or not) the forecasts.

\subsection{The Dynamic Endogeneity KIM}

A more general specification of the score-driven KIM is the Dynamic Endogeneity Kinetic Ising Model (DyEnKIM), where we assume that each parameter  $J$ and $h$ has its own specific time-varying factorization. Going back to Eq. \ref{eq:transprob}, we now impose the following structure to the parameters:
\begin{align}
    \beta = 1 \nonumber \\
    J_{ij}(t) &= \beta_{diag}(t) J_{ij} \delta_{ij} + \beta_{off}(t) J_{ij} (1 - \delta_{ij}) \nonumber \\
    h_i(t) &= \beta_h(t) (h_i + h_0(t)) 
\end{align}
where $\delta_{ij}$ is the Kronecker symbol which is 1 if $i=j$ and 0 otherwise and we will call $\boldsymbol{\beta}(t) = (\beta_{diag}, \beta_{off}, \beta_h )$ in the following.

With this choice we want to be able to discriminate between different components of the observed system dynamics: one associated with the idiosyncratic properties of variable $i$ ($\beta_h$), with  general trends ($h_0$), with autocorrelations ($\beta_{diag}$), and finally with lagged cross-correlations among variables ($\beta_{off}$). In this formulation each of these time-varying parameters $\beta$ measures the relative importance of one term over the others in the generation of the data, highlighting periods of higher endogeneity of the dynamics (when correlations have higher importance) rather than periods where the dynamics is more idiosyncratic or exogenously driven. We report a consistency analysis for the DyEnKIM in the SI, where we show that even under model misspecification this approach correctly separates the different components of the dynamics and captures their relative importance.

\subsection{Role of non stationarity in neural data}

As a first example of the application of the DyEnKIM, we consider the firing dynamics of a set of neurons. Inferring the network of connections between neurons by observing the correlated dynamics of firing has received a lot of attention in the last two decades \cite{cocco2009neuronal, schneidman2006weak} and the KIM has been extensively used for this purpose \cite{hertz2010inferring, zeng2011network, hoang2019network}. The underlying idea is that the (lagged) correlation in the firing of two time series suggests the existence of a physical connection between the two corresponding neurons.

However, as pointed out in \cite{tyrcha2013effect}, correlated behavior can also be generated by the fact that neurons are subject to a common non-stationary input, for example driven by the external environment. Disentangling the contributions to correlations coming from external drivers from those coming from genuine interactions is critical to reliably identify the network structure between neurons.

To this end \cite{tyrcha2013effect} proposes an inferential method to achieve this result by considering a KIM with a time dependent external field $h_i(t)$ representing the contribution of the external stimuli and of all the non recorded neurons to the activity of neuron $i$ at time $t$. However the inference method requires many "trials" or repetitions of the experiment, under the strong methodological assumption that all the repetitions are obtained under identical conditions, an hypothesis that might be difficult to control in such type of complex experiments.

We now show that DyEnKIM can be used for this purpose {\it on a single experiment}. We use the data of \cite{tkavcik2014searching} obtained from a multichannel experiment recording firing patterns of $160$ salamander retina neurons, stimulated by a film clip of a swimming fish. The $20$s experiment is sampled with time binning of $20$ms, corresponding to $T=944$ and we considered the $N=40$ most active neurons. Finally the experiment is repeated $297$ times.

\begin{figure}[t]
    \centering
    \includegraphics[width=.75\linewidth]{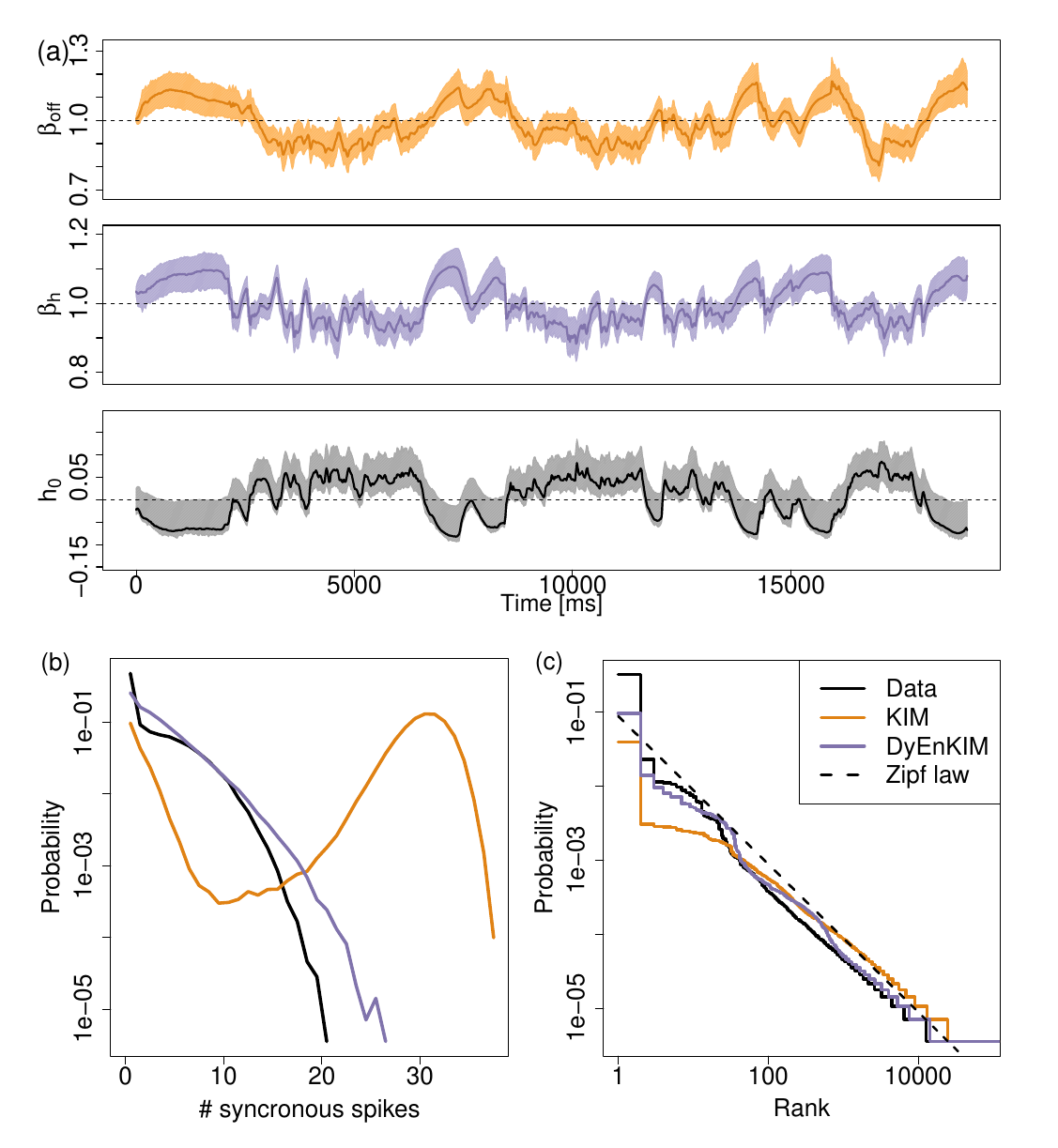}
    \caption{(a) Filtered values of $\beta_{off}$, $\beta_h$, and $h_0(t)$ for salamander retina data. The continuous line is the mean value across the $297$ experiments and the 90\% confidence interval  (i.e. $268/297$ of the filtered values stay within the bands). (b) Estimated probability density function of the number of synchronous spikes. (c) Zipf plot of the frequency of observed patterns. In (b-c) the probability densities are obtained as average across the experiments, but a small variability is observed when considering individual experiments.}
    \label{fig:neu_dye}
\end{figure}

The DyEnKIM of Eq. 4 is estimated and for each experiment we perform an LM test. We find that while for $\beta_{off}(t)$, $\beta_h(t)$, and $h_0(t)$ we reject the null hypothesis of constant parameter in $99.3\%$, $76.8\%$, and $100\%$ of the experiments respectively, this percentage drops to $43.1\%$ for $\beta_{diag}(t)$. For this reason we consider a simplified model where $\beta_{diag}(t)$ is constant\footnote{The following results are essentially unchanged when considering a time varying $\beta_{diag}(t)$.}. Fig. \ref{fig:neu_dye}a shows the temporal dynamics of the three filtered parameters. Since we are able to filter the dynamics for each experiment, in the figure we show the mean and the 90\% confidence interval. It is evident that the three parameters show significant variations, likely in response to the external stimulus provided by the film clip and by unobserved neurons.

In order to evaluate how well our model describes the empirical data we consider two statistics: (i) the distribution of the number of synchronous (i.e., within the same time bin) spikes and (ii) the Zipf plot, obtained as the rank plot of the frequency of each spiking pattern. Both quantities depend on the many body synchronous correlations among spins, thus are not automatically explained by KIM-type models which fit the pairwise correlations. As a benchmark model we consider a constant parameter KIM estimated on the whole dataset.  In Fig. \ref{fig:neu_dye}b-c we show these statistics. We observe that the DyEnKIM reproduces both quite well, while the constant parameter KIM largely fails in describing the distribution of the number of synchronous spikes and in predicting the frequency of the most frequent patterns (rank between $2$ and $\sim 100$) where the underestimation is up to an order of magnitude. We also considered a sparse version of the KIM obtaining similar results (see SI).

The above results are very interesting because they show that a pairwise dynamic interaction model is able to reproduce higher-order correlations, {\it if one takes into account the time varying dynamics of the global interactions} (see also \cite{schneidman2006weak} for the static Ising model). It is important to stress once more that, while an approach as in Ref. \cite{tyrcha2013effect} requires many experiments and the strong methodological assumption that these are identical realizations of the same process, our method to measure time-varying interactions can be performed on a single experiment. Incidentally, one can then use the estimation to test whether the different experiments are statistically equivalent by comparing the estimates across replicas. Moreover our model has only three time dependent scalars, while the model of \cite{tyrcha2013effect} requires a time dependent field for each of the $N$ neurons, thus the latter is highly parametrized with a modeled dynamics strongly constrained by the data.

\subsection{Disentangling endogenous and exogenous price dynamics}


As a second application of the  DyEnKIM we consider the problem of quantifying the contribution to stock price changes due to exogenous events (e.g. news, announcements) and to endogenous feedbacks. A vast literature \cite{filimonov2012quantifying,filimonov2015apparent,hardiman2013critical,hardiman2014branching,rambaldi2015modeling,rambaldi2018detection,wheatley2019endo} has tackled this point, but almost invariably this has been done by assuming that the relation between price and external drivers, as well as those driving the internal feedback, is constant in time. The DyEnKIM allows us to test this hypothesis, by considering time varying parameters whose dynamic can be filtered from data. Understanding the role of  exogenous or endogenous drivers in market volatility  is very important, also to devise possible policy measures able to avoid their occurrences and DyEnKIM, being  able to identify them in real time, could provide valuable tools for market monitoring.

For this application we focus on two events that caused huge turmoil in the stock markets at the intraday level. The first one is the May 6, 2010  Flash Crash, when a seemingly unjustifiable sudden drop in the price of E-mini S\&P 500 futures contracts caused all major stock indices to plummet in a matter of a few minutes, recovering most of the lost value when circuit breakers came into place. Multiple explanations of what happened have been offered by a large number of academics, regulators and practitioners: responsibility has been attributed to careless algorithmic trading \cite{securities2010findings}, deteriorated market liquidity which quickly vanished when price volatility increased \cite{easley2011microstructure}, market fragmentation \cite{madhavan2012exchange, menkveld2019flash}, predatory trading strategies by high-frequency traders \cite{kirilenko2017flash, aquilina2020quantifying}.

The second event we analyze is the announcement following the Federal Open Market Committee (FOMC) meeting of July 31, 2019. In this meeting the Federal Reserve operated its first interest rate cut in over a decade, the last one dating back to the 2008 financial crisis, encountering mixed reactions in both the news and the markets. In particular an answer to a question in the Q\&A press conference by the Fed Chairman Powell has been highlighted by news agencies, when being asked whether further cuts in the future meetings were an option, he answered ``we're thinking of it essentially as a midcycle adjustment to policy'' \cite{powell2019press}. This answer triggered turmoil in the equity markets, with all major indices dropping around $2\%$ in a few minutes.

\begin{figure}[t]
    \centering
    \includegraphics[width=.9\linewidth]{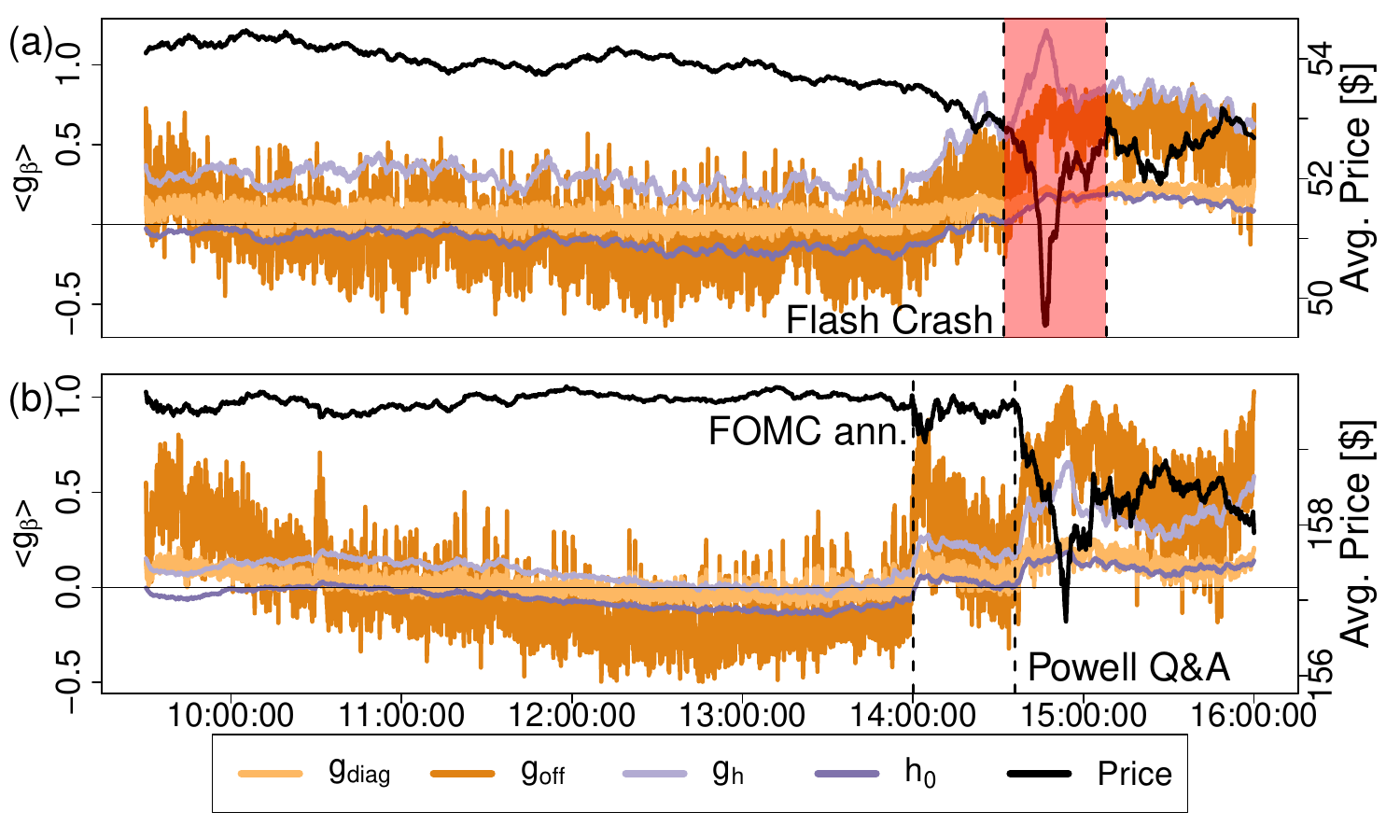}
    \caption{Values of 
    $\langle g_{diag} \rangle (t)$, $\langle g_{off} \rangle (t)$, $\langle g_{h} \rangle (t)$ and $h_0(t)$ during the day of May 6, 2010 Flash Crash (a) and during the day of FOMC announcement on July 31, 2019 (b).
    The black lines are the the average midprice across the S\&P100 stocks. 
    The red area in the top panel highlights the time window (14:32:00 to 15:08:00 EST) where the Flash Crash takes place. 
    }
    \label{fig:fc_dye}
\end{figure}

Like in the previous section, we construct our dataset for both events taking price movements for the then S\&P100-indexed stocks at the 5 seconds time scale and constructing the associated price activity time series. Differently from the previous example, here we apply the DyEnKIM methodology to study variations in the relative importance of different sets of parameters as events unfold. In this case the LM test rejects the null of constant parameters for all $\beta$s and all datasets. To better interpret the results we introduce the value of the components of the effective fields $g_i(t)$, each related to one of the time-varying parameters

\begin{align*}
g_i(t) &= g_{i, diag}(t) + g_{i, off}(t) + g_{i, h}(t) \\
g_{i, diag}(t) &= \beta_{diag}(t) J_{ii} s_i(t) \\
g_{i, off}(t) &= \beta_{off}(t) \sum_j J_{ij} s_j(t) \\
g_{i, h}(t) &= \beta_h(t) (h_i + h_0(t))
\end{align*}
which we then average at each time across all indices $i$, obtaining the quantities $\langle g_{diag} \rangle (t)$ and so on. 


Since the model is applied to price activity, which can be thought of as a proxy of high-frequency volatility \cite{filimonov2012quantifying,hardiman2013critical}, the financial interpretation of these time-varying parameters relates to volatility clustering in the case of $\beta_{diag}$, to volatility spillovers for $\beta_{off}$, to higher or lower market-wise volatility for $h_0$ and the relevance of exogenous effects is given by $\beta_h$.
Thus the $\langle g_\cdot \rangle(t)$ quantities can be intuitively related to what the explained sum of squares means for linear regression models, in the sense that the more a $\langle g_\cdot \rangle (t)$ is far from $0$ relative to others the more the data are affected at time $t$ by that subset of parameters and the corresponding variable. We choose to show these quantities as a simple way of assessing the relevance of the components, a problem that is not easily solved in this kind of models. 


The top panel of Figure \ref{fig:fc_dye} shows the components of the fields during the Flash Crash of May 6, 2010. Here the parameters show a very significant variation around the crash, with a large increase
of $\langle g_h\rangle$
in the 45 minutes preceding the crash together with a similar increase of the endogeneity
field $\langle g_{diag}\rangle$ and $\langle g_{off}\rangle$
during the event, which then stay large until market close. This indicates that the turmoil induced by the Flash Crash reverberated for the remainder of the trading hours, even after the prices had recovered at pre-crash levels. The intraday pattern is overshadowed by the effect of the crash, but the picture at the beginning of the day is similar to normal trading days (see SI).
These results indicate an exogenous increase in activity before the crash, which is accompanied by the endogenous mechanism of volatility spillovers between stocks, as evidenced by  large value of $\langle g_{off}\rangle$  during and after the Flash Crash. In conclusion our analysis indicates that both exogenous and endogenous drivers were important for the onset of Flash Crash.



In the bottom panel of Figure \ref{fig:fc_dye} we show the values of 
the effective fields
on July 31, 2019. The FOMC announcement went public at 14:00:00 EST and is followed by a press conference at 14:30:00 EST, with a Q\&A starting at around 14:36:00 EST. Again we see that the usual intraday pattern is interrupted by the news, which however, differently from the Flash Crash, was a scheduled event. This difference leads to the complete absence of any sort of ``unusual'' effect in the earlier hours of the day, as typically analysts provide forecasts regarding these announcements in the previous days and this information is already incorporated in the prices. What then happens is that, if the news does not meet market expectations, a correction in prices will occur as soon as the information is made public, leading to higher market volatility in the minutes and hours following the announcement \cite{chulia2010asymmetric, hautsch2011impact}. In this specific case, forecasts were mixed between a 0.25\% and a 0.50\% interest rates cut scenario.

The published announcement at 14:00 EST mostly matched these forecasts, with the FOMC lowering the interest target rate by 0.25\%, and we indeed see that the price levels are not particularly affected by the news. However a transient increase in volatility, and in particular the endogenous components, can still be observed in the few minutes following the announcement, quickly returning to average levels. It is interesting to see the reaction to the press conference held 30 minutes after the release, and in particular to the answers the Chairman of the Fed Jerome H. Powell gives to journalists in the Q\&A. As soon as the Q\&A starts, around 14:36 EST, prices begin to plummet in response to the Chairman's answers, possibly reacting to the statement that this interest rates cut was only intended as a ``midcycle adjustment to policy" rather than as the first of a series. Expectations of further rates cuts in the later months of the year could be a reason for this adjustment in the prices when these forecasts are not met, as usually lower interest rates push the stock prices up. We see however that this unexpected event causes a behavior in the estimated time-varying parameters resembling what we have seen in the Flash Crash, albeit the endogenous components are even more significant here. 

\section{Discussion}

We have applied the score-driven methodology to extend the Kinetic Ising Model to a time-varying parameters formulation, introducing two new models for complex systems: the Dynamical Noise Kinetic Ising Model (DyNoKIM) and the Dynamic Endogeneity Kinetic Ising Model (DyEnKIM). We showed that the DyNoKIM, characterized by a time-varying noise level parameter $\beta(t)$, has a clear utility in forecasting applications, as the Area Under the ROC Curve can be showed to be a growing function of $\beta(t)$, while the DyEnKIM can be used to discriminate between endogenous and exogenous effects in the evolution of a multivariate time series. 

We then provided example applications of the two models. We successfully employed the DyNoKIM to quantify the real-time forecasting accuracy of stock price activities in the US stock market, as well as the real-time link prediction accuracy in a temporal social network. The result, largely matching the predictions from theory and simulations, is a methodological breakthrough for the real-world application of time-varying parameter models of complex systems, opening to the possibility of implementing real-time indicators quantifying the accuracy of model-based predictions. 

We have then applied the DyEnKIM to model a population of salamander retina neurons and describe the high-frequency volatility of US stocks in proximity of extreme events such as the Flash Crash of May 6, 2010 or around scheduled announcements as the FOMC report of July 31, 2019. We designed the DyEnKIM to disentangle the effects of interactions from the ones of exogenous sources on the observed collective dynamics, a task that is typically non-trivial but nonetheless fundamental in the modeling of complexity. Our results show that this distinction can be made regardless of the underlying system, providing a detailed description and insight on the dynamics, and most importantly without requiring multiple controlled experiments, as is common practice in previous applications of the KIM on neuron populations, thus opening to the adoption of the model in contexts where running repeated experiments is costly or impossible. 

In conclusion, the Score-Driven KIM poses the foundations for a new modeling paradigm in complex systems. We foresee several relevant extensions such as the modeling of non binary data, for example extending to a Potts-like model \cite{binder1981static}, or to non-Markovian settings. The key advantages provided by the score-driven methodology in terms of ease of estimation and minimization of model misspecification errors open to the implementation of more accurate and versatile models, interesting a wide range of disciplines that look to describe and unravel complexity from empirical observations.

\section{Materials and Methods}

\subsection*{Model inference}

The KIM static parameters $\Theta$ are inferred via Maximum Likelihood Estimation using a known Mean Field technique \cite{mezard2011exact} or, when this is not possible, via standard Gradient Descent methods. Given $\Theta$ we estimate $w, B, A$ by performing a targeted estimation \cite{francq2011merits} through ADAM stochastic Gradient Descent \cite{kingma2014adam}. Targeted estimation, which is common in observation-driven models such as the GARCH \cite{bollerslev1986generalized}, first fits the mean value of the time-varying parameter $\langle f \rangle = w/(1-B)$ and then fits the $(w,B,A)$ parameters keeping this ratio constant. This procedure significantly reduces the estimation time and produces accurate estimates in our simulations. Further details on the process can be found in the SI.

A Lagrange Multiplier (LM) test \cite{calvori2017testing} is used to reject the hypothesis of constant parameters. The test statistic of the LM test can be written as the Explained Sum of Squares (ESS) of the auxiliary linear regression
\begin{equation}
    \mathbf{1} = c_w \nabla^{0}_{t} + c_{A} \mathcal{S}_{(t-1)}^0 \nabla_{t}^{0}
\end{equation}
where $\nabla^{0}_{t}$ is the time $t$ score under the null hypothesis that $f(t) = w \; \forall \, t$, $\mathcal{S}_{t}^0$ is the time $t$ rescaled score (i.e. $\mathcal{I}^{-1/2}(t) \nabla_t$) under the null, the constants $c_w$ and $c_A$ are estimated by standard linear regression methods and the resulting LM test statistic is distributed as a $\chi^2$ random variable with one degree of freedom. If the null is rejected, the hypothesis that $\beta$ is time varying is a valid alternative and we can proceed to estimate the score-driven dynamics parameters. In the DyEnKIM, having multiple time-varying parameters, we test each parameter against two null hypotheses, one where all parameters are constant and one where all other parameters are score-driven, applying FDR correction for multiple tests. All tests on models presented here reject the null with $p < 0.001$.

\subsection*{Data}
US stock prices data provided by LOBSTER academic data - powered by NASDAQ OMX. The data consists of the reconstructed Limit Order Book (LOB) for each US stock with timestamps at millisecond precision. We take the mid-price ({\it i.e.} the average between the best ask and the best bid prices in the LOB) as a real-time proxy of the price, as done in \cite{rambaldi2015modeling}. Press reports about the analyzed market events can be found on financial media outlets. FOMC meeting reports are publicly available at {\it federalreserve.gov}. The salamander retina neuron data has been collected by Prof. Michael J. Berry II and made publicly available at \textit{doi:10.15479/AT:ISTA:61}. It consists of measurements from 160 salamander retina ganglion cells collected through a multi-electrode array. The cells are responding to a light stimulus in the form of a 20 s naturalistic movie and the experiment is repeated 297 times. The electrical signal has been preprocessed to obtain a binary time series for each neuron with time resolution of 20 ms, identifying time intervals where the neuron has produced at least one spike with a 1, and 0 otherwise. From the public dataset we selected the 40 neurons with highest average spike rate over the 297 repeats of the experiment.

The data describing situations of face to face proximity between individuals in the workplace, is provided by the SocioPatterns \cite{sociopatternswebsite} collaboration. 
It was collected, over a period of two weeks, in one of the two office buildings of the Institut National de Veille Sanitaire (InVS), located in Saint Maurice near Paris, France. Two thirds of the total staff agreed to participate to the data collection. They were asked to wear a sensor on their chest, that allow exchange of radio packets only when the persons are facing each other at a range closer than 1.5 m. By design, any contact that lasted at least 20 seconds was recorded with a probability higher than 99\%. In our temporal network application, we associate a node to each individual, and assign a link between two workers if they face each other at a distance less than $1.5$ meters. We then consider only the subset of the $100$ most active links in each day.

\section*{Acknowledgements}
Part of this work has been supported by the European Integrated Infrastructure for Social Mining and Big Data Analytics (SoBigData++, Grant Agreement \#871042). C.C. acknowledges support from the Swiss National Science Foundation grant \#200021\_182659.

\newpage
\bibliographystyle{unsrt}%
\bibliography{biblio.bib}%

\begin{thebibliography}{100}

\bibitem{bar2002general}
Yaneer Bar-Yam.
\newblock General features of complex systems.
\newblock {\em Encyclopedia of Life Support Systems (EOLSS), UNESCO, EOLSS
  Publishers, Oxford, UK}, 1, 2002.

\bibitem{challet2016trader}
Damien Challet, R{\'e}my Chicheportiche, Mehdi Lallouache, and Serge
  Kassibrakis.
\newblock Trader lead-lag networks and order flow prediction.
\newblock {\em Available at SSRN 2839312}, 2016.

\bibitem{lillo2015news}
Fabrizio Lillo, Salvatore Miccich{\`e}, Michele Tumminello, Jyrki Piilo, and
  Rosario~N Mantegna.
\newblock How news affects the trading behaviour of different categories of
  investors in a financial market.
\newblock {\em Quantitative Finance}, 15(2):213--229, 2015.

\bibitem{schmitz2017predator}
Oswald Schmitz.
\newblock Predator and prey functional traits: understanding the adaptive
  machinery driving predator--prey interactions.
\newblock {\em F1000Research}, 6, 2017.

\bibitem{tavoni2017functional}
Gaia Tavoni, Ulisse Ferrari, Francesco~P Battaglia, Simona Cocco, and R{\'e}mi
  Monasson.
\newblock Functional coupling networks inferred from prefrontal cortex activity
  show experience-related effective plasticity.
\newblock {\em Network Neuroscience}, 1(3):275--301, 2017.

\bibitem{crisanti1988dynamics}
A~Crisanti and Haim Sompolinsky.
\newblock Dynamics of spin systems with randomly asymmetric bonds: Ising spins
  and glauber dynamics.
\newblock {\em Physical Review A}, 37(12):4865, 1988.

\bibitem{ackley1985learning}
David~H Ackley, Geoffrey~E Hinton, and Terrence~J Sejnowski.
\newblock A learning algorithm for boltzmann machines.
\newblock {\em Cognitive science}, 9(1):147--169, 1985.

\bibitem{cox1981statistical}
David~R Cox, Gudmundur Gudmundsson, Georg Lindgren, Lennart Bondesson, Erik
  Harsaae, Petter Laake, Katarina Juselius, and Steffen~L Lauritzen.
\newblock Statistical analysis of time series: Some recent developments [with
  discussion and reply].
\newblock {\em Scandinavian Journal of Statistics}, pages 93--115, 1981.

\bibitem{creal2013generalized}
Drew Creal, Siem~Jan Koopman, and Andr{\'e} Lucas.
\newblock Generalized autoregressive score models with applications.
\newblock {\em Journal of Applied Econometrics}, 28(5):777--795, 2013.

\bibitem{harvey_2013}
Andrew~C. Harvey.
\newblock {\em Dynamic Models for Volatility and Heavy Tails: With Applications
  to Financial and Economic Time Series}.
\newblock Econometric Society Monographs. Cambridge University Press, 2013.

\bibitem{blasques2015information}
Francisco Blasques, Siem~Jan Koopman, and Andre Lucas.
\newblock Information-theoretic optimality of observation-driven time series
  models for continuous responses.
\newblock {\em Biometrika}, 102(2):325--343, 2015.

\bibitem{bernardi2019switching}
Mauro Bernardi and Leopoldo Catania.
\newblock Switching generalized autoregressive score copula models with
  application to systemic risk.
\newblock {\em Journal of Applied Econometrics}, 34(1):43--65, 2019.

\bibitem{di2019score}
Domenico Di~Gangi, Giacomo Bormetti, and Fabrizio Lillo.
\newblock Score-driven exponential random graphs: A new class of time-varying
  parameter models for dynamical networks.
\newblock {\em arXiv preprint arXiv:1905.10806}, 2019.

\bibitem{nelson}
Daniel~B. Nelson.
\newblock Filtering and forecasting with misspecified arch models i: Getting
  the right variance with the wrong model.
\newblock {\em Journal of Econometrics}, 52:61--90, 1992.

\bibitem{derrida1987exactly}
Bernard Derrida, Elizabeth Gardner, and Anne Zippelius.
\newblock An exactly solvable asymmetric neural network model.
\newblock {\em EPL (Europhysics Letters)}, 4(2):167, 1987.

\bibitem{kirkpatrick1978infinite}
Scott Kirkpatrick and David Sherrington.
\newblock Infinite-ranged models of spin-glasses.
\newblock {\em Physical Review B}, 17(11):4384, 1978.

\bibitem{edwards1975theory}
Samuel~Frederick Edwards and Phil~W Anderson.
\newblock Theory of spin glasses.
\newblock {\em Journal of Physics F: Metal Physics}, 5(5):965, 1975.

\bibitem{jaynes1957information}
Edwin~T Jaynes.
\newblock Information theory and statistical mechanics.
\newblock {\em Physical review}, 106(4):620, 1957.

\bibitem{schneidman2006weak}
Elad Schneidman, Michael~J Berry, Ronen Segev, and William Bialek.
\newblock Weak pairwise correlations imply strongly correlated network states
  in a neural population.
\newblock {\em Nature}, 440(7087):1007--1012, 2006.

\bibitem{marre2009prediction}
Olivier Marre, Sami El~Boustani, Yves Fr{\'e}gnac, and Alain Destexhe.
\newblock Prediction of spatiotemporal patterns of neural activity from
  pairwise correlations.
\newblock {\em Physical review letters}, 102(13):138101, 2009.

\bibitem{cocco2017functional}
Simona Cocco, R{\'e}mi Monasson, Lorenzo Posani, and Gaia Tavoni.
\newblock Functional networks from inverse modeling of neural population
  activity.
\newblock {\em Current Opinion in Systems Biology}, 3:103--110, 2017.

\bibitem{nghiem2018maximum}
Trang-Anh Nghiem, Bartosz Telenczuk, Olivier Marre, Alain Destexhe, and Ulisse
  Ferrari.
\newblock Maximum-entropy models reveal the excitatory and inhibitory
  correlation structures in cortical neuronal activity.
\newblock {\em Physical Review E}, 98(1):012402, 2018.

\bibitem{ferrari2018separating}
Ulisse Ferrari, St{\'e}phane Deny, Matthew Chalk, Ga{\v{s}}per Tka{\v{c}}ik,
  Olivier Marre, and Thierry Mora.
\newblock Separating intrinsic interactions from extrinsic correlations in a
  network of sensory neurons.
\newblock {\em Physical Review E}, 98(4):042410, 2018.

\bibitem{tanaka1977model}
Seiji Tanaka and Harold~A Scheraga.
\newblock Model of protein folding: incorporation of a one-dimensional
  short-range (ising) model into a three-dimensional model.
\newblock {\em Proceedings of the National Academy of Sciences},
  74(4):1320--1323, 1977.

\bibitem{imparato2007ising}
A~Imparato, A~Pelizzola, and M~Zamparo.
\newblock Ising-like model for protein mechanical unfolding.
\newblock {\em Physical review letters}, 98(14):148102, 2007.

\bibitem{agliari2011thermodynamic}
Elena Agliari, Adriano Barra, Francesco Guerra, and Francesco Moauro.
\newblock A thermodynamic perspective of immune capabilities.
\newblock {\em Journal of theoretical biology}, 287:48--63, 2011.

\bibitem{bornholdt2001expectation}
Stefan Bornholdt.
\newblock Expectation bubbles in a spin model of markets: Intermittency from
  frustration across scales.
\newblock {\em International Journal of Modern Physics C}, 12(05):667--674,
  2001.

\bibitem{bouchaud2013}
Jean-Philippe Bouchaud.
\newblock Crises and collective socio-economic phenomena: Simple models and
  challenges.
\newblock {\em Journal of Statistical Physics}, 151(3):567--606, May 2013.

\bibitem{sornette2014physics}
Didier Sornette.
\newblock Physics and financial economics (1776--2014): puzzles, ising and
  agent-based models.
\newblock {\em Reports on progress in physics}, 77(6):062001, 2014.

\bibitem{campajola2020unveiling}
Carlo Campajola, Fabrizio Lillo, and Daniele Tantari.
\newblock Unveiling the relation between herding and liquidity with trader
  lead-lag networks.
\newblock {\em Quantitative Finance}, 20(11):1765--1778, 2020.

\bibitem{lecun2015deep}
Yann LeCun, Yoshua Bengio, and Geoffrey Hinton.
\newblock Deep learning.
\newblock {\em Nature}, 521(7553):436, 2015.

\bibitem{hornik1989multilayer}
Kurt Hornik, Maxwell Stinchcombe, and Halbert White.
\newblock Multilayer feedforward networks are universal approximators.
\newblock {\em Neural networks}, 2(5):359--366, 1989.

\bibitem{decelle2015inference}
Aur{\'e}lien Decelle and Pan Zhang.
\newblock Inference of the sparse kinetic ising model using the decimation
  method.
\newblock {\em Physical Review E}, 91(5):052136, 2015.

\bibitem{campajola2019inference}
Carlo Campajola, Fabrizio Lillo, and Daniele Tantari.
\newblock Inference of the kinetic ising model with heterogeneous missing data.
\newblock {\em Physical Review E}, 99(6):062138, 2019.

\bibitem{hanley1982meaning}
James~A Hanley and Barbara~J McNeil.
\newblock The meaning and use of the area under a receiver operating
  characteristic (roc) curve.
\newblock {\em Radiology}, 143(1):29--36, 1982.

\bibitem{bradley1997use}
Andrew~P Bradley.
\newblock The use of the area under the roc curve in the evaluation of machine
  learning algorithms.
\newblock {\em Pattern recognition}, 30(7):1145--1159, 1997.

\bibitem{penney1993coupled}
RW~Penney, ACC Coolen, and D~Sherrington.
\newblock Coupled dynamics of fast spins and slow interactions in neural
  networks and spin systems.
\newblock {\em Journal of Physics A: Mathematical and General}, 26(15):3681,
  1993.

\bibitem{beck2003superstatistics}
Christian Beck and Ezechiel~GD Cohen.
\newblock Superstatistics.
\newblock {\em Physica A: Statistical mechanics and its applications},
  322:267--275, 2003.

\bibitem{beck2005time}
Christian Beck, Ezechiel~GD Cohen, and Harry~L Swinney.
\newblock From time series to superstatistics.
\newblock {\em Physical Review E}, 72(5):056133, 2005.

\bibitem{mezard2011exact}
Marc M{\'e}zard and J~Sakellariou.
\newblock Exact mean-field inference in asymmetric kinetic ising systems.
\newblock {\em Journal of Statistical Mechanics: Theory and Experiment},
  2011(07):L07001, 2011.

\bibitem{ait2011ultra}
Yacine A{\"\i}t-Sahalia, Per~A Mykland, and Lan Zhang.
\newblock Ultra high frequency volatility estimation with dependent
  microstructure noise.
\newblock {\em Journal of Econometrics}, 160(1):160--175, 2011.

\bibitem{filimonov2012quantifying}
Vladimir Filimonov and Didier Sornette.
\newblock Quantifying reflexivity in financial markets: Toward a prediction of
  flash crashes.
\newblock {\em Physical Review E}, 85(5):056108, 2012.

\bibitem{hardiman2013critical}
Stephen~J Hardiman, Nicolas Bercot, and Jean-Philippe Bouchaud.
\newblock Critical reflexivity in financial markets: a hawkes process analysis.
\newblock {\em The European Physical Journal B}, 86(10):442, 2013.

\bibitem{filimonov2015apparent}
Vladimir Filimonov and Didier Sornette.
\newblock Apparent criticality and calibration issues in the hawkes
  self-excited point process model: application to high-frequency financial
  data.
\newblock {\em Quantitative Finance}, 15(8):1293--1314, 2015.

\bibitem{hardiman2014branching}
Stephen~J Hardiman and Jean-Philippe Bouchaud.
\newblock Branching-ratio approximation for the self-exciting hawkes process.
\newblock {\em Physical Review E}, 90(6):062807, 2014.

\bibitem{wheatley2019endo}
Spencer Wheatley, Alexander Wehrli, and Didier Sornette.
\newblock The endo--exo problem in high frequency financial price fluctuations
  and rejecting criticality.
\newblock {\em Quantitative Finance}, 19(7):1165--1178, 2019.

\bibitem{rambaldi2015modeling}
Marcello Rambaldi, Paris Pennesi, and Fabrizio Lillo.
\newblock Modeling foreign exchange market activity around macroeconomic news:
  Hawkes-process approach.
\newblock {\em Physical Review E}, 91(1):012819, 2015.

\bibitem{rambaldi2018detection}
Marcello Rambaldi, Vladimir Filimonov, and Fabrizio Lillo.
\newblock Detection of intensity bursts using hawkes processes: An application
  to high-frequency financial data.
\newblock {\em Physical Review E}, 97(3):032318, 2018.

\bibitem{calvori2017testing}
Francesco Calvori, Drew Creal, Siem~Jan Koopman, and Andr{\'e} Lucas.
\newblock Testing for parameter instability across different modeling
  frameworks.
\newblock {\em Journal of Financial Econometrics}, 15(2):223--246, 2017.

\bibitem{newman2006structure}
Mark~Ed Newman, Albert-L{\'a}szl{\'o}~Ed Barab{\'a}si, and Duncan~J Watts.
\newblock {\em The structure and dynamics of networks.}
\newblock Princeton university press, 2006.

\bibitem{cohen2010complex}
Reuven Cohen and Shlomo Havlin.
\newblock {\em Complex networks: structure, robustness and function}.
\newblock Cambridge university press, 2010.

\bibitem{barabasi2013network}
Albert-L{\'a}szl{\'o} Barab{\'a}si.
\newblock Network science.
\newblock {\em Philosophical Transactions of the Royal Society A: Mathematical,
  Physical and Engineering Sciences}, 371(1987):20120375, 2013.

\bibitem{newman2018networks}
Mark Newman.
\newblock {\em Networks}.
\newblock Oxford university press, 2018.

\bibitem{gao2013understanding}
Song Gao, Yaoli Wang, Yong Gao, and Yu~Liu.
\newblock Understanding urban traffic-flow characteristics: a rethinking of
  betweenness centrality.
\newblock {\em Environment and Planning B: Planning and Design},
  40(1):135--153, 2013.

\bibitem{fagiolo2013international}
Giorgio Fagiolo and Marina Mastrorillo.
\newblock International migration network: Topology and modeling.
\newblock {\em Physical Review E}, 88(1):012812, 2013.

\bibitem{draief2010epidemics}
Moez Draief and Laurent Massoulie.
\newblock {\em Epidemics and rumours in complex networks}, volume 369.
\newblock Cambridge University Press Cambridge, 2010.

\bibitem{bhattacharya2008international}
Kunal Bhattacharya, Gautam Mukherjee, Jari Saram{\"a}ki, Kimmo Kaski, and
  Subhrangshu~S Manna.
\newblock The international trade network: weighted network analysis and
  modelling.
\newblock {\em Journal of Statistical Mechanics: Theory and Experiment},
  2008(02):P02002, 2008.

\bibitem{gai2011complexity}
Prasanna Gai, Andrew Haldane, and Sujit Kapadia.
\newblock Complexity, concentration and contagion.
\newblock {\em Journal of Monetary Economics}, 58(5):453--470, 2011.

\bibitem{cimini2015systemic}
Giulio Cimini, Tiziano Squartini, Diego Garlaschelli, and Andrea Gabrielli.
\newblock Systemic risk analysis on reconstructed economic and financial
  networks.
\newblock {\em Scientific reports}, 5(1):1--12, 2015.

\bibitem{holme2012temporal}
Petter Holme and Jari Saram{\"a}ki.
\newblock Temporal networks.
\newblock {\em Physics reports}, 519(3):97--125, 2012.

\bibitem{sewell2015latent}
Daniel~K Sewell and Yuguo Chen.
\newblock Latent space models for dynamic networks.
\newblock {\em Journal of the American Statistical Association},
  110(512):1646--1657, 2015.

\bibitem{sewell2018simultaneous}
Daniel~K Sewell.
\newblock Simultaneous and temporal autoregressive network models.
\newblock {\em Network Science}, 6(2):204--231, 2018.

\bibitem{mazzarisi2020dynamic}
Piero Mazzarisi, Paolo Barucca, Fabrizio Lillo, and Daniele Tantari.
\newblock A dynamic network model with persistent links and node-specific
  latent variables, with an application to the interbank market.
\newblock {\em European Journal of Operational Research}, 281(1):50--65, 2020.

\bibitem{Hanneke_tergm_10.1214/09-EJS548}
Steve Hanneke, Wenjie Fu, and Eric~P. Xing.
\newblock {Discrete temporal models of social networks}.
\newblock {\em Electronic Journal of Statistics}, 4(none):585 -- 605, 2010.

\bibitem{holl1981}
Paul~W. Holland and Samuel Leinhardt.
\newblock An exponential family of probability distributions for directed
  graphs.
\newblock {\em Journal of the American Statistical Association},
  76(373):33--50, 1981.

\bibitem{wang2015link}
Peng Wang, BaoWen Xu, YuRong Wu, and XiaoYu Zhou.
\newblock Link prediction in social networks: the state-of-the-art.
\newblock {\em Science China Information Sciences}, 58(1):1--38, 2015.

\bibitem{martinez2016survey}
V{\'\i}ctor Mart{\'\i}nez, Fernando Berzal, and Juan-Carlos Cubero.
\newblock A survey of link prediction in complex networks.
\newblock {\em ACM computing surveys (CSUR)}, 49(4):1--33, 2016.

\bibitem{genois2015data}
Mathieu G{\'e}nois, Christian~L Vestergaard, Julie Fournet, Andr{\'e} Panisson,
  Isabelle Bonmarin, and Alain Barrat.
\newblock Data on face-to-face contacts in an office building suggest a
  low-cost vaccination strategy based on community linkers.
\newblock {\em Network Science}, 3(3):326--347, 2015.

\bibitem{sociopatternswebsite}
{SocioPatterns Research Collaboration}.
\newblock {http://www.sociopatterns.org/}, 2008.

\bibitem{cocco2009neuronal}
Simona Cocco, Stanislas Leibler, and R{\'e}mi Monasson.
\newblock Neuronal couplings between retinal ganglion cells inferred by
  efficient inverse statistical physics methods.
\newblock {\em Proceedings of the National Academy of Sciences},
  106(33):14058--14062, 2009.

\bibitem{hertz2010inferring}
John~A Hertz, Yasser Roudi, Andreas Thorning, Joanna Tyrcha, Erik Aurell, and
  Hong-Li Zeng.
\newblock Inferring network connectivity using kinetic ising models.
\newblock {\em BMC neuroscience}, 11(1):1--2, 2010.

\bibitem{zeng2011network}
Hong-Li Zeng, Erik Aurell, Mikko Alava, and Hamed Mahmoudi.
\newblock Network inference using asynchronously updated kinetic ising model.
\newblock {\em Physical Review E}, 83(4):041135, 2011.

\bibitem{hoang2019network}
Danh-Tai Hoang, Juyong Song, Vipul Periwal, and Junghyo Jo.
\newblock Network inference in stochastic systems from neurons to currencies:
  Improved performance at small sample size.
\newblock {\em Physical Review E}, 99(2):023311, 2019.

\bibitem{tyrcha2013effect}
Joanna Tyrcha, Yasser Roudi, Matteo Marsili, and John Hertz.
\newblock The effect of nonstationarity on models inferred from neural data.
\newblock {\em Journal of Statistical Mechanics: Theory and Experiment},
  2013(03):P03005, 2013.

\bibitem{tkavcik2014searching}
Ga{\v{s}}per Tka{\v{c}}ik, Olivier Marre, Dario Amodei, Elad Schneidman,
  William Bialek, and Michael~J Berry~II.
\newblock Searching for collective behavior in a large network of sensory
  neurons.
\newblock {\em PLoS Comput Biol}, 10(1):e1003408, 2014.

\bibitem{securities2010findings}
US~Securities \&~Exchange Commission and Commodity Futures~Trading Commission.
\newblock Findings regarding the market events of may 6, 2010.
\newblock {\em Washington DC}, 2010.

\bibitem{easley2011microstructure}
David Easley, Marcos M~Lopez De~Prado, and Maureen O'Hara.
\newblock The microstructure of the ``flash crash": flow toxicity, liquidity
  crashes, and the probability of informed trading.
\newblock {\em The Journal of Portfolio Management}, 37(2):118--128, 2011.

\bibitem{madhavan2012exchange}
Ananth Madhavan.
\newblock Exchange-traded funds, market structure, and the flash crash.
\newblock {\em Financial Analysts Journal}, 68(4):20--35, 2012.

\bibitem{menkveld2019flash}
Albert~J Menkveld and Bart~Zhou Yueshen.
\newblock The flash crash: A cautionary tale about highly fragmented markets.
\newblock {\em Management Science}, 65(10):4470--4488, 2019.

\bibitem{kirilenko2017flash}
Andrei Kirilenko, Albert~S Kyle, Mehrdad Samadi, and Tugkan Tuzun.
\newblock The flash crash: High-frequency trading in an electronic market.
\newblock {\em The Journal of Finance}, 72(3):967--998, 2017.

\bibitem{aquilina2020quantifying}
Matteo Aquilina, Eric Budish, and Peter O'Neill.
\newblock Quantifying the high-frequency trading ``arms race": A simple new
  methodology and estimates.
\newblock {\em United Kingdom Financial Conduct Authority Occasional Paper},
  2020.

\bibitem{powell2019press}
Jerome Powell.
\newblock Transcript of chair powell's press conference.
\newblock {\em Federal Open Market Committee}, July 31, 2019.

\bibitem{chulia2010asymmetric}
Helena Chuli{\'a}, Martin Martens, and Dick van Dijk.
\newblock Asymmetric effects of federal funds target rate changes on s\&p100
  stock returns, volatilities and correlations.
\newblock {\em Journal of Banking \& Finance}, 34(4):834--839, 2010.

\bibitem{hautsch2011impact}
Nikolaus Hautsch, Dieter Hess, and David Veredas.
\newblock The impact of macroeconomic news on quote adjustments, noise, and
  informational volatility.
\newblock {\em Journal of Banking \& Finance}, 35(10):2733--2746, 2011.

\bibitem{binder1981static}
K~Binder.
\newblock Static and dynamic critical phenomena of the two-dimensional q-state
  potts model.
\newblock {\em Journal of Statistical Physics}, 24(1):69--86, 1981.

\bibitem{francq2011merits}
Christian Francq, Lajos Horvath, and Jean-Michel Zako{\"\i}an.
\newblock Merits and drawbacks of variance targeting in garch models.
\newblock {\em Journal of Financial Econometrics}, 9(4):619--656, 2011.

\bibitem{kingma2014adam}
Diederik~P Kingma and Jimmy Ba.
\newblock Adam: A method for stochastic optimization.
\newblock {\em arXiv preprint arXiv:1412.6980}, 2014.

\bibitem{bollerslev1986generalized}
Tim Bollerslev.
\newblock Generalized autoregressive conditional heteroskedasticity.
\newblock {\em Journal of econometrics}, 31(3):307--327, 1986.

\bibitem{ising1925beitrag}
Ernst Ising.
\newblock Beitrag zur theorie des ferromagnetismus.
\newblock {\em Zeitschrift f{\"u}r Physik A Hadrons and Nuclei},
  31(1):253--258, 1925.

\bibitem{onsager1944crystal}
Lars Onsager.
\newblock Crystal statistics. i. a two-dimensional model with an order-disorder
  transition.
\newblock {\em Physical Review}, 65(3-4):117, 1944.

\bibitem{hopfield1982neural}
John~J Hopfield.
\newblock Neural networks and physical systems with emergent collective
  computational abilities.
\newblock {\em Proceedings of the national academy of sciences},
  79(8):2554--2558, 1982.

\bibitem{shannon1948mathematical}
Claude~E Shannon.
\newblock A mathematical theory of communication.
\newblock {\em Bell system technical journal}, 27(3):379--423, 1948.

\bibitem{coolen2001statistical}
ACC Coolen.
\newblock Statistical mechanics of recurrent neural networks i: Statics.
\newblock {\em Handbook of biological physics}, 4:531--596, 2001.

\bibitem{coolen2001statistical2}
ACC Coolen.
\newblock Statistical mechanics of recurrent neural networks iidynamics.
\newblock {\em Handbook of biological physics}, 4:619--684, 2001.

\bibitem{dunn2013learning}
Benjamin Dunn and Yasser Roudi.
\newblock Learning and inference in a nonequilibrium ising model with hidden
  nodes.
\newblock {\em Physical Review E}, 87(2):022127, 2013.

\bibitem{jacobs1978discrete}
Patricia~A Jacobs and Peter~AW Lewis.
\newblock Discrete time series generated by mixtures. {III}. autoregressive
  processes ({DAR} (p)).
\newblock Technical report, Naval Postgraduate School Monterey Calif, 1978.

\bibitem{campajola2021equivalence}
Carlo Campajola, Fabrizio Lillo, Piero Mazzarisi, and Daniele Tantari.
\newblock On the equivalence between the kinetic ising model and discrete
  autoregressive processes.
\newblock {\em Journal of Statistical Mechanics: Theory and Experiment},
  2021(3):033412, 2021.

\bibitem{tucci1995time}
Marco~P Tucci.
\newblock Time-varying parameters: a critical introduction.
\newblock {\em Structural Change and Economic Dynamics}, 6(2):237--260, 1995.

\bibitem{koopman2016predicting}
Siem~Jan Koopman, Andre Lucas, and Marcel Scharth.
\newblock Predicting time-varying parameters with parameter-driven and
  observation-driven models.
\newblock {\em Review of Economics and Statistics}, 98(1):97--110, 2016.

\bibitem{tauchen1983price}
George~E Tauchen and Mark Pitts.
\newblock The price variability-volume relationship on speculative markets.
\newblock {\em Econometrica: Journal of the Econometric Society}, pages
  485--505, 1983.

\bibitem{shephard2005stochastic}
Neil Shephard.
\newblock {\em Stochastic volatility: selected readings}.
\newblock Oxford University Press on Demand, 2005.

\bibitem{bauwens2004stochastic}
Luc Bauwens and David Veredas.
\newblock The stochastic conditional duration model: a latent variable model
  for the analysis of financial durations.
\newblock {\em Journal of econometrics}, 119(2):381--412, 2004.

\bibitem{hafner2012dynamic}
Christian~M Hafner and Hans Manner.
\newblock Dynamic stochastic copula models: Estimation, inference and
  applications.
\newblock {\em Journal of Applied Econometrics}, 27(2):269--295, 2012.

\bibitem{blasques2017finite}
Francisco Blasques, Andre Lucas, and Andries van Vlodrop.
\newblock Finite sample optimality of score-driven volatility models.
\newblock {\em Tinbergen Institute Discussion Paper}, 17-111/III, 2017.

\bibitem{nelson1992filtering}
Daniel~B Nelson.
\newblock Filtering and forecasting with misspecified arch models i: Getting
  the right variance with the wrong model.
\newblock {\em Journal of econometrics}, 52(1-2):61--90, 1992.

\bibitem{roudi2011mean}
Yasser Roudi and John Hertz.
\newblock Mean field theory for nonequilibrium network reconstruction.
\newblock {\em Physical review letters}, 106(4):048702, 2011.

\end{thebibliography}

\newpage
\begin{appendices}
\numberwithin{equation}{section}
\numberwithin{figure}{section}
\section{Further information on the Score-Driven KIM}

\subsection{Additional information on Kinetic Ising Models}

Spin systems have been analyzed by physicists since the early 20th century, mostly as models to understand the microscopical foundations of magnetism. A spin is indeed a proxy for an atomic magnetic moment, i.e. the torque the atom is subject to when immersed in a magnetic field. The first spin model is the celebrated 1D Ising Model \cite{ising1925beitrag}, which mathematically abstracts the problem to an infinite chain of binary variables ${s}$ (the spins) that can be either in an ``up'' or ``down'' state. These perceive the local magnetic field generated by their nearest neighbors as well as any external magnetic field and tend to ``align'' (i.e. match the state) or ``disalign'' (i.e. go in the opposite state) with the net field they sense, based on the value of a parameter $J$. The model was intended to verify the hypothesis that thermal properties of ferromagnetic materials can arise from microscopic interactions between their atoms, but failed to do so because no net magnetic field would be observed at equilibrium. However, it was later shown \cite{onsager1944crystal} that the failure was not due to the mechanism, but to the oversimplification of taking a 1-dimensional system: in fact, if one takes a 2D lattice instead of a 1D chain, this extremely simple model qualitatively reproduces the macroscopic thermal properties of ferromagnets. The success of the Ising Model has led to its extension and refinement to describe exotic materials such as spin glasses \cite{kirkpatrick1978infinite}, and its fascinating ability to describe macroscopic properties determined by microscopic coordination posed the foundations to many quantitative models of complex systems, with examples of successful Ising-like models for protein and DNA chains \cite{tanaka1977model}, neurons \cite{hopfield1982neural} and financial markets \cite{bouchaud2013}.

The appeal of Ising Models comes in part from the fact that they belong to the class of \textit{Maximum Entropy} models, as introduced by \cite{jaynes1957information}. The principle states that, given a set of constrained quantities from available observations - such as sample averages - a probability distribution that maximizes Shannon's entropy \cite{shannon1948mathematical} subject to the constraints is the best distribution to describe the observations, as it is the one that makes the least arbitrary assumptions. In particular Ising Models result from Shannon's entropy maximization constraining means and correlations of the spins, thus making them a popular choice to describe systems that can be encoded in binary strings.

The Kinetic Ising Model (KIM) is the out-of-equilibrium version of the Sherrington-Kirkpatrick (SK) spin glass \cite{derrida1987exactly, crisanti1988dynamics}, developed a few years later and proposed as dynamical model for asymmetric neural networks with discrete time and synchronous sampling. The model's transition probability, describing the probability of observing a future configuration $\lbrace s_i(t)\rbrace$ given a current configuration $\lbrace s_t(t-1)\rbrace$, reads

\begin{equation}\label{eq:transprob_app}
p(\lbrace s_i(t) \rbrace \vert \lbrace s_i(t-1) \rbrace, J, h) = \frac{1}{Z(t)} \exp \lbrace \sum_{i,j} J_{ij} s_i(t) s_j(t-1) + \sum_i h_i s_i(t) \rbrace
\end{equation}

Differently from its predecessor, which describes the equilibrium properties of spin glasses, this model describes the dynamics of a system of spins which have asymmetric interactions, namely spin $i$'s effect on spin $j$ is different from spin $j$'s effect on $i$. This difference is incorporated in the structure of the $J$ matrix, which is symmetric in the SK model and asymmetric in the KIM. Having $J_{ij} \neq J_{ji}$ in fact implies that these coefficients can no longer describe a synchronous interaction, as for instance a correlation coefficient, but need to describe an asynchronous one, specifically in this case a lag one interaction.

Typically, in the physics literature, the $J$ elements are assumed to be $iid$ Gaussian random variables, $J_{ij} \sim \mathcal{N}(J_0/N, J_1^2/N)$ and the properties of the model as data generating process are the object of analysis. As shown in \cite{crisanti1988dynamics}, the KIM loses the so-called ``spin glass'' phase of the SK model - a phase in which the system ``freezes'' in a metastable configuration with local order but no global order - and only presents a dynamic phase transition between a paramagnetic phase - where spins do not show preferential alignment - and a ferromagnetic phase - where all spins align in one direction - when the mean of the $J$ elements, $J_0/N$, is greater than $1/N$.

A more complete characterization of the model can be found in the physics literature \cite{crisanti1988dynamics, derrida1987exactly, coolen2001statistical, coolen2001statistical2}, with recent developments contributing to  neuroscience \cite{tyrcha2013effect}, machine learning \cite{dunn2013learning, decelle2015inference, campajola2019inference} and finance \cite{campajola2020unveiling} literatures. As a last remark, the model has been developed in at least another independent strand of literature with the name of Discrete AutoRegressive model (DAR) \cite{jacobs1978discrete}, and an equivalence between these models  has been recently shown in \cite{campajola2021equivalence}.

\subsection{Additional information on score-driven models}

Let us set the stage to better explain score-driven models by briefly reviewing the theory of time-varying parameters models in discrete time. There is a rich literature on the topic, which has been summarized in the review by Tucci \cite{tucci1995time} and more recently by Koopman et al. \cite{koopman2016predicting}. In general, a time-varying parameters model can be written as 

\begin{subequations}
\begin{equation}
    s(t) \sim p(s(t) \vert f(t), \mathcal{S}(t-1), \Phi_1)
\end{equation}
\begin{equation}
    f(t) = \psi(f(t-1), f(t-2), ..., \mathcal{S}(t-1), \epsilon(t), \Phi_2)
\end{equation}
\end{subequations}
where $s(t)$ is a vector of observations sampled from the probability distribution function $p$, $\mathcal{S}(t-1)$ is the set of all observations up to time $t-1$ and $f(t)$ are the parameters which are assumed to be time varying. The dynamics of those parameters can either depend on past observations, on past values of the same parameters, on some external noise $\epsilon(t)$ and on two sets of static parameters $\Phi_1$ and $\Phi_2$. 

If the function $\psi$ only contains past values of the time-varying parameters, a noise term and the static parameters, then the model is called a \textit{parameter-driven} model, whereas if the function $\psi$ can be written as a deterministic function only of past observations and past parameters, it is called an \textit{observation-driven} model \cite{cox1981statistical}.

Examples for parameter-driven models can be found in the financial econometrics literature looking at stochastic volatility models \cite{tauchen1983price, shephard2005stochastic}, which aim at describing the time-varying nature of the volatility ({\it i.e.} the variance) of price variations, as well as other examples \cite{bauwens2004stochastic, hafner2012dynamic}.

Within the observation-driven models, the most celebrated example is the Generalized AutoRegressive Conditional Heteroscedasticity (GARCH) model \cite{bollerslev1986generalized}, where a time series of financial log-returns is modelled using a time-varying volatility parameter depending deterministically on squared observations up to that time and past values of volatilities.

The main advantage of adopting an observation-driven model rather than a parameter-driven one lies in its estimation: having time-varying parameters that only depend on observations through a set of static parameters results in a strong reduction of complexity in writing the likelihood of the model, whereas the calculations for most non-trivial parameter-driven models are typically extremely convoluted and computationally intensive.

Score-driven (or Generalized Autoregressive Score - GAS - models) are a specific class of observation-driven models. Originally introduced by Creal et al. \cite{creal2013generalized} and Harvey \cite{harvey_2013}, they postulate that time-varying parameters depend on observations through the score of the conditional likelihood, that the gradient of its logarithm.

Let us restate Eq. 2 of the main text to provide a more detailed explanation of the score-driven dynamics

\begin{equation}  \label{eq:gasupdaterule_app}
f(t+1) = w + B f(t) + A \mathcal{I}^{-1/2}(t) \nabla_t
\end{equation}
where $w$, $B$ and $A$ are a set of static parameters, $f(t) \in \mathbb{R}^M$ is a vector of time-varying parameters of the model's conditional probability density $p(s(t) \vert f(t), ...)$ with $s(t) \in \mathbb{R}^N$, and $\nabla_t = \frac{\partial \log p (s(t) \vert f(t))}{\partial f(t)}$ is the score. In this generic form, $w$ is a $M$-dimensional vector, while $A$ and $B$ are $M \times M$ matrices. $\mathcal{I}^{-1/2}(t)$ is also a $M\times M$ matrix, introduced to rescale the time $t$ score to account for local convexity, that we choose to be the inverse of the square root of the Fisher information matrix associated with $p(s(t) \vert f(t))$. This is not the only possible choice for this rescaling matrix \cite{creal2013generalized} but in our opinion it is the most intuitive way of rescaling the score (and probably the most common one).

As mentioned, one of the main reasons to choose an observation-driven model is the less challenging estimation, but it is not the only one. Score-driven models in particular have been shown to be optimal
in terms of Kullback-Leibler divergence \cite{blasques2015information, blasques2017finite} in approximating any unknown underlying probability distribution. Given the absence of unobservable noise processes, contrary to parameter-driven models in general, they are able to properly fit unknown parameter dynamics with accuracy that are second only to the data generating process itself. This makes score-driven models the ideal choice whenever prior knowledge about parameters dynamics is scarce.

Finally, the score-driven modelling approach provides access to a simple Lagrange Multiplier statistical test \cite{calvori2017testing}, of the null hypothesis that a given parameter is constant. This is of crucial importance when estimating a model parameters from data, as knowing whether the parameter can be considered static or should be assumed to be time-varying helps in the selection of models that extract more relevant informations from the data and are less prone to overfitting or underfitting problems. We reported details on the Lagrange Multiplier test in the ``Materials and Methods'' section in the main text.

\subsection{Score-Driven models as filters of a misspecified dynamics}

Here we present a simple example of how score-driven models can be used to filter an unknown dynamics of a parameter, without assuming a specific model for its time evolution. We will do it by considering the classical case of a discrete time random walk model with time-varying diffusion coefficient. This type of models is very popular in finance where the (logarithm of the) price follows a random walk and the diffusion rate, termed volatility, represents the risk of the asset. As we will show, under minimal assumptions, such a filter turns out to coincide with the popular GARCH model for volatility.

All this is relatively well known. Indeed, the interpretation of GARCH processes as predictive filters is well described in this statement by Nelson \cite{nelson1992filtering}: ``Note that our use of the term `estimate' corresponds to its use in the filtering literature rather than the statistics literature; that is, an ARCH model with (given) fixed parameters produces `estimates' of the true underlying conditional covariance matrix at each point in time in the same sense that a Kalman filter produces `estimates' of unobserved state variables in a linear system''.

Let us call $s(t)$ the increment of the log-price, $p(t+1)-p(t)$ and consider a  stochastic volatility model 
$$
s(t)=\sigma(t) \epsilon(t)~~~~~~~~~\epsilon(t)\sim {\mathcal N}(0,1). 
$$
i.e. the conditional probability density function of $s(t)$ is
$$
p(s(t)|\sigma(t))=\frac{1}{\sqrt{2\pi \sigma^2(t)}} e^{-\frac{s^2(t)}{2\sigma^2(t)}}
$$

By choosing as time-varying parameter $f(t)=\sigma^2(t)$, the score of the likelihood is 
$$
\frac{\partial\log p(s(t)|f(t))}{\partial f(t)}=-\frac{1}{2\sigma^2(t)}+\frac{s^2(t)}{2\sigma^4(t)}
$$
hence the equation for the evolution of volatility is 
$$
\sigma^2(t+1)=w+B\sigma^2(t)+\frac{A \mathcal{I}^{-1/2}(t)}{2}\left[\frac{s^2(t)-\sigma^2(t)}{\sigma^4(t)}\right]
$$
Thus if $s^2(t)\gg \sigma^2(t)$ ($s^2(t)\ll \sigma^2(t)$), the new $\sigma^2(t+1)$ will be larger (smaller) than $\sigma^2(t)$. This is exactly the mechanism which dynamically adjusts the filtered estimation of volatility taking into account the most recent observation(s).

\begin{figure}[h]
\centering
\includegraphics[width=1\textwidth]{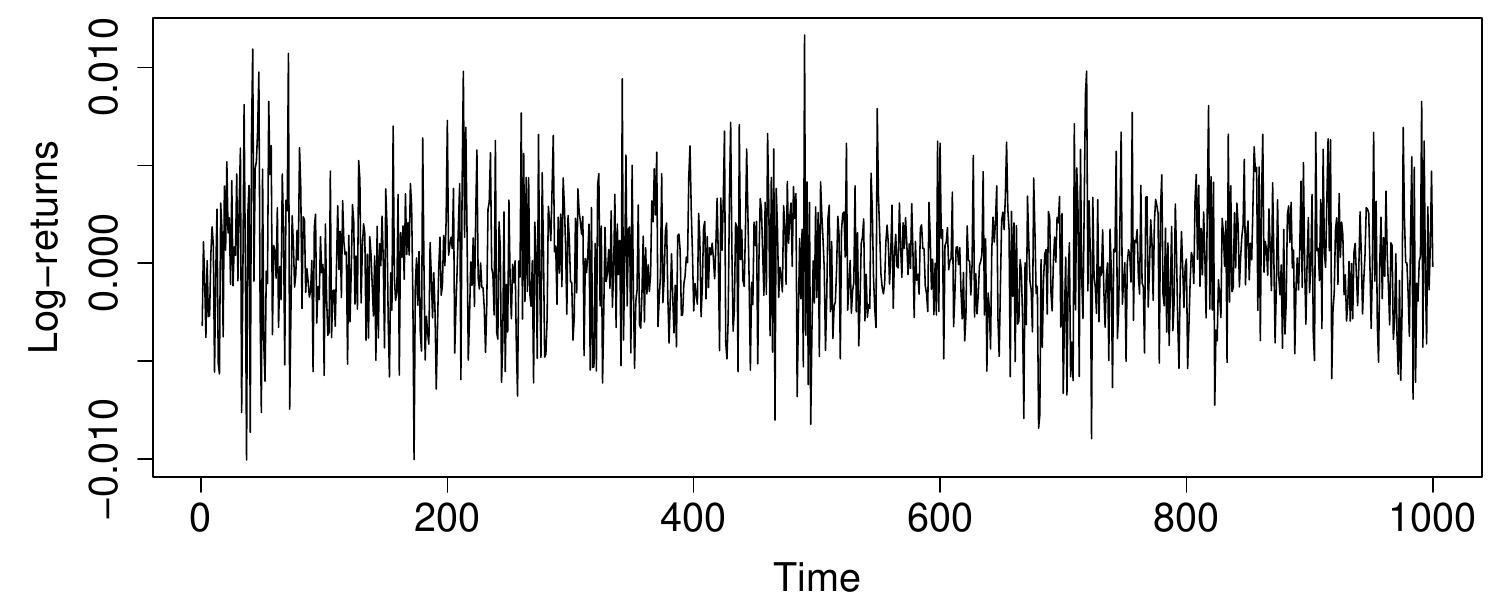}
\caption{Artificially generated time series of length $1000$ from a GARCH(1,1)  model with $w = 10^{-6}$, $\alpha = 0.1$, $\beta = 0.8$.}

\label{fig:garch}
\end{figure}

By choosing $\mathcal{I}(t)$ as the the Fisher information matrix and using  ${\mathbb E}[s^2(t)|\sigma^2(t)]=\sigma^2(t)$, it is
$$
\mathcal{I}(t)\equiv -{\mathbb E}\left[\frac{\partial^2 \log p(s(t)|\sigma(t))}{\partial^2 \sigma^2(t)}\bigg | \sigma^2(t)\right]=-{\mathbb E}\left[\frac{1}{2\sigma^4(t)}-\frac{s^2(t)}{\sigma^6(t)}\bigg | \sigma^2(t) \right]=\frac{1}{2\sigma^4(t)}
$$
 thus
\begin{equation}\label{eq:garch}
\sigma^2(t+1)=w+B\sigma^2(t)+A(s^2(t)-\sigma^2(t))=w+\alpha s^2(t)+\beta \sigma^2(t)
\end{equation}
with $\alpha=A$ and $\beta=B-A$, which coincides with the GARCH model. This model has been originally proposed as a data generating process for describing realistic dynamics of volatility, while here it is derived as a result of Score-Driven modeling. The GARCH model of Eq. \ref{eq:garch} is typically seen as a data generating process for the volatility, and thus the price, of financial assets. This model is routinely estimated from real data and used widely in the financial industry for risk management, portfolio allocation, systemic risk, etc.. Fig. \ref{fig:garch} shows a typical simulated price pattern from a GARCH(1,1) process, displaying fat tails and clustered volatility, as observed in empirical data. Other popular econometric models as Multiplicative Error Model (MEM), Autoregressive Conditional Duration (ACD), Autoregressive Conditional Intensity (ACI) can be cast as special cases of score-driven models.

\begin{figure}[h]
\centering
\includegraphics[width=1\textwidth]{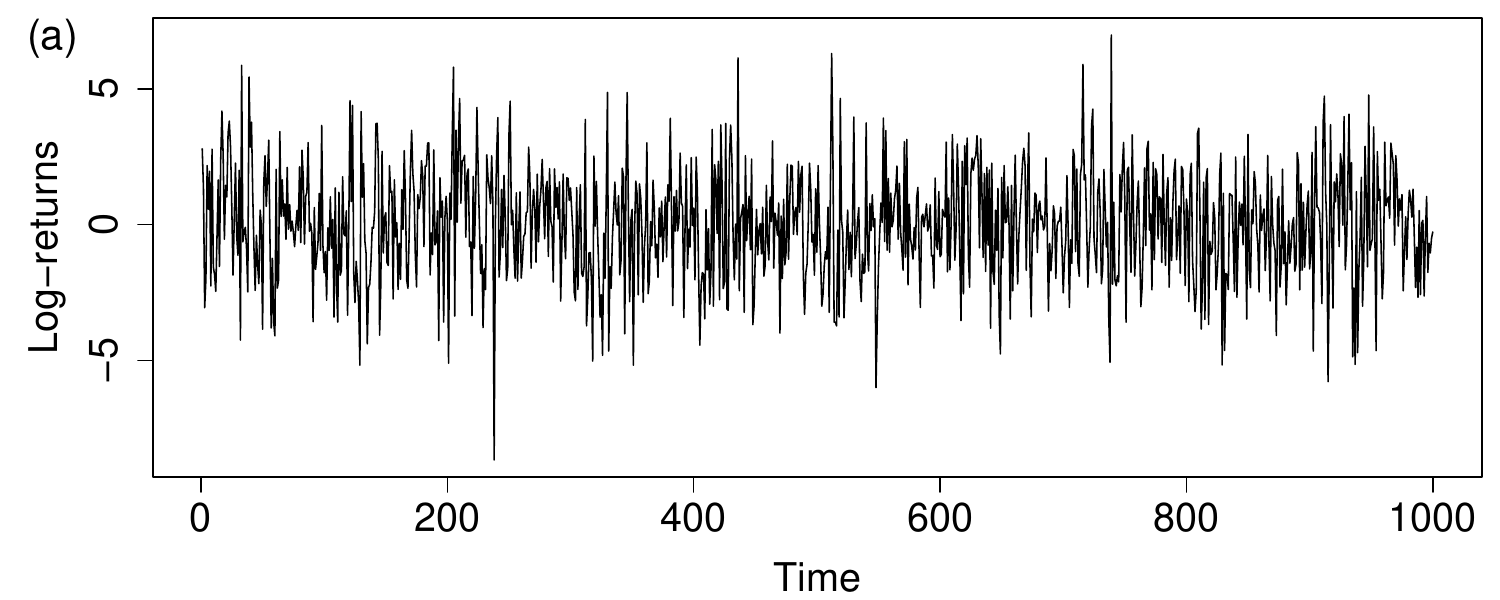} \\
\includegraphics[width=1\textwidth]{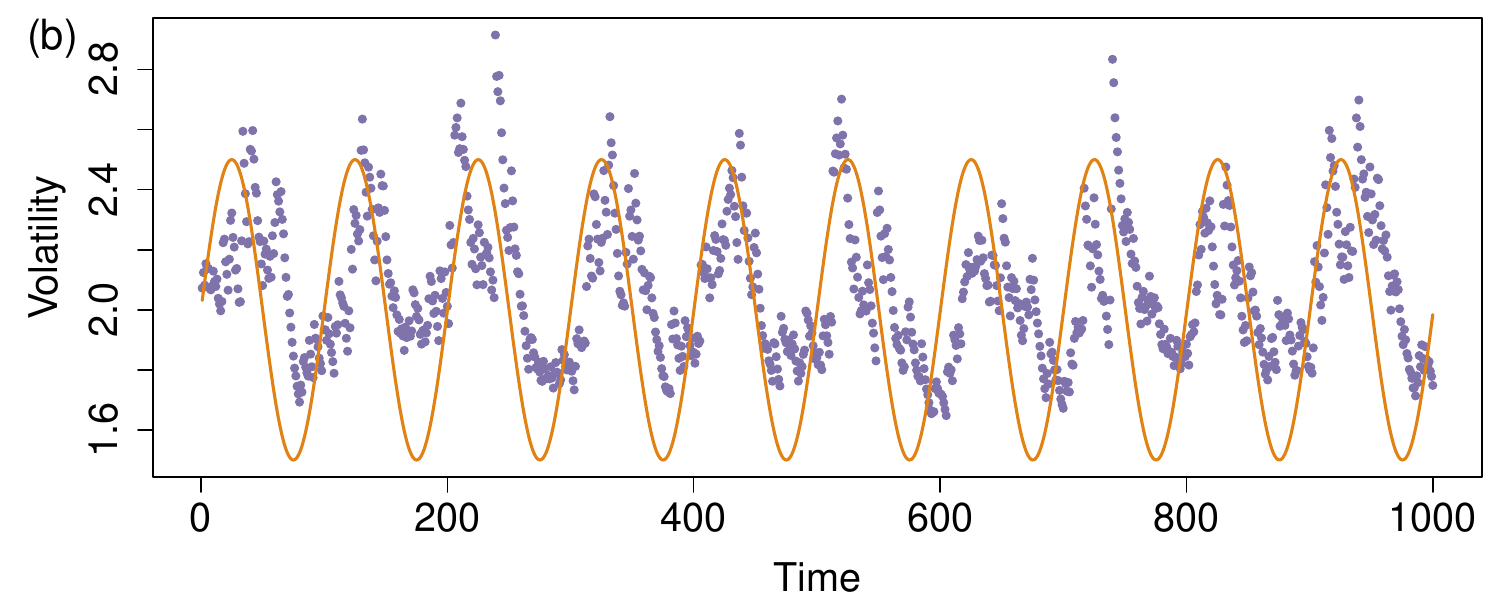}
\caption{(a) Artificially generated time series of returns according to the model of Eq. \ref{eq:sine}. (b) 
Simulated (orange line) and filtered (purple dots) values of $\sigma(t)$. The latter are obtained by fitting a GARCH(1,1) model on the data in the left panel.}
\label{fig:garchfilter}
\end{figure}

However, the main point we want to make here concerns the use of GARCH, and more generally of score-driven models, as {\it filters} of a differently specified dynamics. To show this in practice, we simulate $1000$ price observations from the model
\begin{equation}\label{eq:sine}
s(t)=\sigma(t) \epsilon(t)~~~~~\epsilon(t)\sim {\mathcal N}(0,1)~~~~~\sigma(t)=2+\frac{1}{2}\sin\left(\frac{\pi t}{100}\right)
\end{equation}
The left panel of Fig. \ref{fig:garchfilter} shows the simulated price dynamics. This is clearly {\it not} a GARCH model and the sinuisodal shape can be modified with other deterministic or stochastic processes. Assuming the data generating process of Eq. \ref{eq:sine} is unknown, one can nevertheless fit the GARCH(1,1) model and obtain, beside the static parameters $w$, $\alpha$, and $\beta$, the filtered values of $\sigma(t)$. The outcome of this procedure is shown in the right panel of Fig. \ref{fig:garchfilter} where the red line is the simulated $\sigma(t)$, while the black circles represent the filtered values of $\sigma(t)$.

The example shows how score-driven models can be used to filter the time-varying parameters with unknown dynamics from data. As mentioned in the main text,  Score-driven models have been shown to be an optimal choice among observation-driven models when minimising the Kullback-Leibler divergence to an unknown generating probability distribution \cite{blasques2015information}.

\subsection{Details on the inference method}

As mentioned in the main text our estimation procedure is done in steps, starting by estimating the parameters $\Theta = (J,h)$ of the standard KIM and then running a targeted estimation for the $w$, $B$ and $A$ parameters. In this Appendix we provide some further details about this procedure.

The whole process can be summarized as the maximization of the log-likelihood $\mathcal{L}(\Theta, \beta(t), w, B, A)$ of the model in question, which in the case of the DyNoKIM reads (setting as usual $h_i = 0$ $\forall i$)

\begin{subequations}
\begin{equation}
    \mathcal{L}(\Theta, \beta(t), w, B, A) = \sum_{t=1}^T \left\lbrace \sum_i \left[ \beta(t) \sum_j s_i(t) J_{ij} s_j(t-1) \right] - \log Z(t) \right\rbrace \label{eq:loglik}
\end{equation}
\begin{equation}
    \mathrm{with} \; \; \log \beta(t+1) = w + B \log \beta(t) + A \mathcal{I}^{-1/2} (t) \nabla_t \label{eq:betadynapp}
\end{equation}
\end{subequations}
and the definitions of the various quantities are given in the main text. The log-likelihood shown above has a recursive form, as each term in the sum of Eq. \ref{eq:loglik} depends on $\beta(t)$, which is determined recursively through Eq. \ref{eq:betadynapp} from a starting condition $\beta(1)$. This means that, if one were to maximize $\mathcal{L}$ with respect to all the parameters by applying a standard Gradient Descent method, at each computation of $\mathcal{L}$ and its gradient it would be necessary to compute the recursion, resulting in a slow and computationally cumbersome process. In order to make the estimation quicker we implement our multi-step procedure, relying on existing methods for the estimation of the standard KIM and of observation-driven models.

Our first step consists of maximizing $\mathcal{L}$ with respect to the standard KIM parameters $\Theta$. This is done adopting the Mean Field approach of M\'{e}zard and Sakellariou \cite{mezard2011exact}, which is both fast and accurate in the estimation of fully connected models. We refer the interested readers to the original publication for further details on the method itself. In the specific case of neuron spike data, the Mean Field method fails numerically and we resort to standard Gradient Descent methods. The main reason to detach this step from the optimization of the complete log-likelihood is that $\Theta$ contains a large number of parameters: if one can get an estimate for those without recurring to slow and hard to tune Gradient Descent methods the computational cost of the inference reduces significantly.

Given the values of $\Theta$ obtained in the first step, we then move to the targeted estimation of $w$, $B$ and $A$. This consists in first estimating a target value $\bar{f}$ for the unconditional mean of $f(t)= \log \beta(t)$ and then optimize $w$, $B$ and $A$ maintaining the ratio $w/(1-B) = \bar{f}$ fixed. To estimate $\bar{f}$ we maximize the log-likelihood of Eq. \ref{eq:loglik} temporarily imposing $A=B=0$, hence Eq. \ref{eq:betadynapp} becomes $\log \beta(t) = \bar{f} = \mathrm{const}$. Finally, given this target value we optimize $\mathcal{L}$ with respect to $w$, $B$ and $A$ maintaining the ratio $w/(1-B) = \bar{f}$ fixed and setting $f(1) = \bar{f}$ to start the recursion of Eq. \ref{eq:betadynapp}. During these last two steps we use the ADAptive Momentum (ADAM) \cite{kingma2014adam} Stochastic Gradient Descent method as optimization algorithm, as we found in our case it had better performance with respect to other available methods.

This targeted estimation is not necessary - one could directly estimate $w$, $B$ and $A$ together - but it is a standard procedure in the estimation of observation-driven models like the GARCH \cite{francq2011merits}, as it typically reduces the total number of iterations of gradient descent.

We point out one last remark concerning the indetermination of $\langle \beta \rangle$ in the model of Eq. 3 in the main text (and of $\langle \boldsymbol{\beta} \rangle$ for the DyEnKIM), which is crucial to understand the results of our simulations. The fact that these values cannot be identified is not problematic {\it per se}, but requires caution when comparing models and filtered parameters across different samples, or when comparing estimates with simulations. To avoid misleading results, one needs to enforce the sample mean of the filtered $\beta(t)$ (or of each of the elements of $\boldsymbol{\beta}(t)$ in the DyEnKIM) to be equal to a reference value, which without loss of generality we pick to be $\langle \beta \rangle = 1$. This is easily done by running the estimation and filtering, then measuring $\langle \beta \rangle$ and rescaling $\beta' (t) = \beta(t)/ \langle \beta \rangle$. To leave the model unchanged an opposite rescaling is needed for the parameters $J$ and $h$, each having to be multiplied by $\langle \beta \rangle$ themselves. This transformation does not change the log-likelihood, thus the model parameters are still MLE, but crucially allows to set a reference value for $\beta$ that solves the indetermination.

Given this remark, in all the simulations we show where the data generating process of $\beta(t)$ is misspecified we generate its values making sure that their sample mean is $1$. By doing so we do not lose any generality in our results, as the indetermination needs to be solved for the data generating process too if one wants to obtain meaningful results, and we are able to correctly compare the simulated values of $J$, $h$ and $\beta(t)$ with the ones that are estimated by the score-driven model. Notably, since the model is misspecified, this cannot be achieved during estimation by enforcing the targeted unconditional mean to be equal to $1$, as the score in that case is not a martingale difference and thus the unconditional mean of the score-driven parameter is ill-defined itself, as shown by Creal et al. \cite{creal2013generalized}.

\subsection{Derivation of the theoretical AUC}

Here we expand on the derivation of the theoretical Area Under the ROC Curve shown in Fig. 1 in the main text. A ROC curve is a set of points ($FPR(\alpha), TPR(\alpha)$), with $\alpha \in [0,1]$ being a free parameter determining the minimum value of $p(s_i(t) = +1 \vert s(t-1);\beta, \Theta)$ which is considered to predict $\hat{s}_i(t) = 1$. If the prediction $\hat{s}_i(t)$ matches the realization $s_i(t)$ then the classification is identified as a True Positive (or Negative, if $p < \alpha$), otherwise it is identified as a False Positive (Negative). The True Positive Rate (TPR) is the ratio of True Positives to the total number of realized Positives, that is True Positives plus False Negatives. Similarly the False Positive Rate (FPR) is the ratio of False Positives to the total number of realized Negatives. Summarizing

\begin{align*}
    TPR = \frac{TP}{TP + FN}; ~~~~~~~~~~
    FPR = \frac{FP}{FP + TN}
\end{align*}

We can explicitly derive the analytical form of the theoretical Area Under the Curve, that is the area that lies below the set of points $(FPR(\alpha), TPR(\alpha))$, assuming the data generating process is well specified and performing some assumptions on the distribution of the model parameters. As a reminder, a classifier having $AUC=0.5$ is called an \textit{uninformed classifier}, meaning it makes predictions statistically indistinguishable from random guessing, while values of $AUC$ greater than $0.5$ are a sign of good forecasting capability.

Following the definition of TPR and FPR one can compute their expected values

\begin{subequations}\label{tprfpr:main}
\begin{equation}
TPR_\phi(\alpha, \beta) = \frac{1}{Z^+_\phi(\beta)} \int_{g_i : p^+ > \alpha} dg_i \phi(g) p^+(\beta,g_i) \label{tprfpr:tpr}
\end{equation}
\begin{equation}
FPR_\phi(\alpha, \beta) = \frac{1}{Z^-_\phi(\beta)} \int_{g_i : p^+ > \alpha} dg_i \phi(g) p^-(\beta,g_i) \label{tprfpr:fpr}
\end{equation}
where $Z^\pm_\phi (\beta) = p(s_i = \pm 1)$ is a normalization function, $\phi(g)$ is the unconditional distribution of the effective fields $g_i$ and we have abbreviated the probability of sampling a positive or negative value as
\begin{equation}
p^\pm(\beta,g_i) = \frac{e^{ \pm \beta g_i}}{2 \cosh (\beta g_i)} \nonumber
\end{equation}
\end{subequations}

The definition of the theoretical AUC then reads as

\begin{equation*}
AUC_\phi(\beta) = \int_1^0 TPR_\phi(\alpha,\beta) \frac{\partial FPR_\phi(\alpha, \beta )}{\partial \alpha } d\alpha 
\end{equation*}

that is the area below the set of points $(FPR(\alpha), TPR(\alpha))$. The lower limit to the integration in Eqs. \ref{tprfpr:main} is $g_{min} : p^+(g_{min}) = \alpha$, which is found to be

\begin{equation*}
    g_{min}(\alpha,\beta) = \frac{1}{2\beta} \log \frac{\alpha}{1 - \alpha}
\end{equation*}

Then applying the partial derivative to the definition of FPR it follows that

\begin{equation*}
\frac{\partial FPR}{\partial \alpha} = - \frac{1}{Z^-_\phi(\beta)} \frac{\partial g_{min}}{\partial \alpha} \phi(g_{min}) (1 - \alpha)
\end{equation*}

where we have substituted $p^- (\beta, g_{min}) = 1 - \alpha$. Plugging all the above results in the definition of $AUC_\phi$ we then find

\begin{equation} \label{eq:theorauc}
AUC_\phi(\beta) = \frac{1}{Z^+_\phi(\beta)Z^-_\phi(\beta)} \int_0^1 d\alpha \left[ \int_{g_{min}(\alpha,\beta)}^{+\infty} dg \phi(g) \frac{e^{\beta g}}{2 \cosh \beta g} \right] \left[ \frac{1}{2 \alpha \beta} \phi(g_{min}(\alpha,\beta)) \right]
\end{equation}

From an operational perspective $\phi(g)$ is the distribution that the effective fields show cross-sectionally across the whole sample, that is $g_i(t) \sim \phi(g)$ $\forall i,t$, but it can also be calculated by giving a prior distribution to the static parameters of the model, $\Theta = (J,h,b)$. Finding this distribution can be useful to provide an easier and more accurate evaluation of the expected AUC of a forecast at a given $\beta$ value, as it provides a bridge from the model parameters to the $AUC(\beta)$ we derived in Eq. \ref{eq:theorauc} and shown in Fig. 1 in the main text.

Let us assume, as is standard in the literature \cite{crisanti1988dynamics, roudi2011mean,mezard2011exact}, that the parameters $\Theta$ are structured in such a way that

\begin{align*}
J_{ij} &\overset{iid}{\sim} \mathcal{N}(J_0/N, J_1^2/N - J_0^2/N^2) \\
h_i &\overset{iid}{\sim} \mathcal{N}(h_0, h_1^2)
\end{align*}

If that is the case then the distribution of $g_i(t)$ is itself a Gaussian, as $g_i(t)$ is now a sum of independent Gaussian random variables $J_{ij}$ and $h_i$ with random coefficients $s_j(t)$. Let us also define two average operators: the average $\langle \cdot \rangle$ over the distribution $p$, also called the \textit{thermal} average (which, the system being ergodic, coincides with a time average for $T \rightarrow \infty$), and the average $\overline{\; \cdot \;}$ over the distribution of parameters, also known as the \textit{disorder} average. Following M\'{e}zard and Sakellariou \cite{mezard2011exact} we can then find the unconditional mean of $s_i$ which reads

\begin{equation}\label{eq:uncmean_i}
    m_i = \langle s_i(t) \rangle = \left\langle \tanh \left[ \beta g_i(t) \right] \right\rangle
\end{equation}

where we have substituted the conditional mean value of $s_i(t)$ inside the brackets. This depends from the distribution of $g_i(t)$: assuming stationarity and calling $g_i^0 = \langle g_i(t) \rangle$ and $\Delta_i^2 = \langle g_i^2(t) \rangle - \langle g_i(t) \rangle^2$ we find that they are

\begin{subequations}
\begin{equation}\label{eq:gmean}
    g_i^0 = \langle \sum_j J_{ij} s_j(t) + h_i \rangle = \sum_j J_{ij} m_j + h_i
\end{equation}
\begin{equation}\label{eq:gvariance}
    \Delta_i^2 = \left\langle \left( \sum_j J_{ij} s_j(t) + h_i \right)^2 \right\rangle - \left\langle \sum_j J_{ij}s_j(t) + h_i \right\rangle^2 = \sum_{j,k} J_{ij} J_{ik} \left[ \left\langle s_j(t) s_k(t) \right\rangle - m_jm_k \right]
\end{equation}
\end{subequations}

In Eq. \ref{eq:gvariance} spins $s_j(t)$ and $s_k(t)$ are mutually conditionally independent under distribution $p$: this means that the only surviving terms are the ones for $j=k$, and thus we find

\begin{equation}\label{eq:gvariance2}
    \Delta_i^2 = \sum_j J_{ij}^2 (1 - m_j^2)
\end{equation}

Having determined the value of the mean and variance of the effective field of spin $i$ we can now proceed to average over the disorder and find the unconditional distribution of effective fields at any time and for any spin, $\phi(g)$. First we realize that the average of Eq. \ref{eq:uncmean_i} can now be substituted by a Gaussian integral

\begin{equation}
    m_i = \int Dx \tanh \left[\beta \left(g_i^0 + x \Delta_i \right) \right]
\end{equation}

where $Dx$ is a Gaussian measure of variable $x \sim \mathcal{N}(0,1)$. Then we can see that the unconditional mean of the fields distribution $\phi(g)$ is

\begin{equation}
   g_0 = \overline{\langle g_i(t) \rangle} = \overline{\sum_j J_{ij} m_j + h_i}
\end{equation}

Given the above results and the definition of $J$, the dependency between $J_{ij}$ and $m_j$ vanishes like $O(1/N)$, which means that the two can be averaged over the disorder separately in the limit $N \rightarrow \infty$. 
This results in the following expression for the unconditional mean of $g_i(t)$

\begin{equation}\label{eq:gmean_res}
    g_0 = J_0 \overline{m_j} + h_0 = J_0 m + h_0
\end{equation}

where

$$
m = \overline{m_i} = \overline{\int Dx \tanh \left[ \beta (g_i + x\Delta_i )\right]}
$$

both the integral and the average here are of difficult solution and results have been provided by Crisanti and Sompolinsky \cite{crisanti1988dynamics}: they show that in the limit $N \rightarrow \infty$ and with $h_i=0$ $\forall i$ the system can be in one of two phases, a paramagnetic phase where $m = 0$ if $\beta$ is smaller than a critical threshold $\beta_c(J_0)$ and $J_0 < 1$, and a ferromagnetic phase where $m \neq 0$ otherwise. In the following we report results for simulations in the paramagnetic phase, as the inference is not possible in the ferromagnetic phase. To give better intuition let us consider the integral above in the limit $\beta \rightarrow 0$: then we can expand the hyperbolic tangent around $0$ to find (since $x$ has zero mean)

\begin{equation}\label{eq:smallbeta}
m \approx \overline{\beta g_i} = \overline{\beta \left(\sum_j J_{ij} m_j + h_0 \right)} = \beta (J_0 m + h_0)
\end{equation}

which in turn leads to an approximated solution for $g_0$ in the limit $\beta \rightarrow 0$

$$
g_0 \approx h_0 \left( \frac{\beta J_0}{1 - \beta J_0} + 1 \right)
$$

Moving on to the variance of $g$ the calculation is straightforward. Adding the mean over the disorder to Eq. \ref{eq:gvariance} we find

\begin{align}\label{eq:gvar_res}
    g_1^2 &= \overline{\left \langle \left[ \sum_j J_{ij} s_j(t) + h_i \right]^2 \right \rangle} - \overline{\left \langle \sum_j J_{ij} s_j(t) + h_i \right \rangle}^2 = \nonumber \\
    &= \overline{\sum_{j} J_{ij}^2 + h_i^2 + 2 h_i \sum_j J_{ij} m_j} - \overline{\sum_j J_{ij} m_j + h_i}^2 = \nonumber \\
    &= J_1^2 + h_1^2 - J_0^2 m^2 
\end{align}

Equations \ref{eq:gmean_res} and \ref{eq:gvar_res} can then be used to calculate, given the parameters of the distribution generating $\Theta$, the values of $g_0$ and $g_1$ that are to be plugged in the distribution $\phi(g)$ of Eq. \ref{eq:theorauc}

\begin{figure}[h]
    \centering
    \includegraphics[width=1\textwidth]{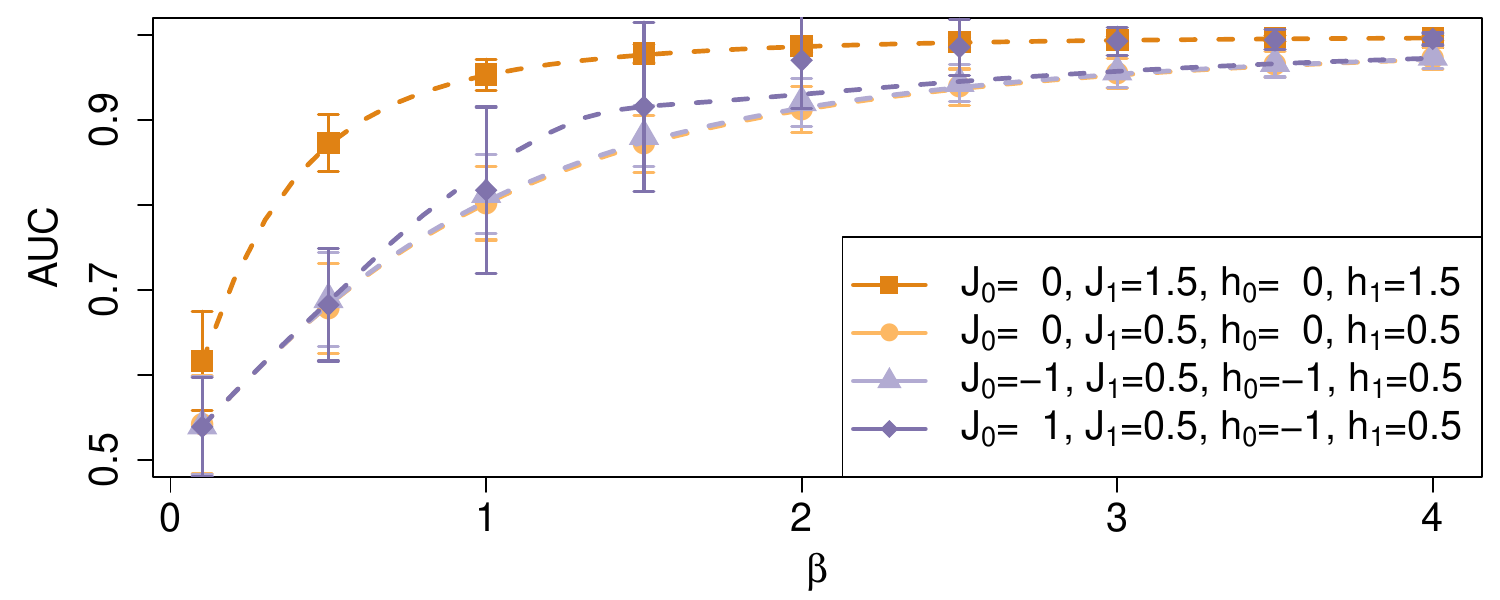}
    \caption{Comparison between the AUC estimated on data simulated from a Kinetic Ising Model and the theoretically derived AUC with Gaussian distribution of the $J$ and $h$ parameters, varying $\beta$ and the hyperparameters $J_0$, $J_1$, $h_0$ and $h_1$. Plot points report average simulated values for a given $\beta$ with error bars at $\pm 1$ standard deviation, dashed lines report theoretical values predicted by Eq. \ref{eq:theorauc}.}
    \label{fig:gaussauc_sim}
\end{figure}

We simulated a Kinetic Ising Model with $N=100$ spins for $T=2000$ time steps at different constant values of $\beta$ and then measured the AUC of predictions assuming the parameters are known. In Fig. \ref{fig:gaussauc_sim} we report a comparison between these simulated values and the theoretical ones provided by Eq. \ref{eq:theorauc} varying $\beta$ and the hyperparameters $J_0$, $J_1$, $h_0$ and $h_1$ in the Gaussian setting we just discussed and adopting the expansion for $\beta \rightarrow 0$. We see that the approximation for small $\beta$ of Eq. \ref{eq:smallbeta} does not affect the accuracy of the theoretical prediction for larger values of $\beta$ and that the mean is correctly captured by Eq. \ref{eq:theorauc}. The only exception to this is found for $\beta > 1$ and $J_0 = 1$, which according to the literature is close to the line of the ferromagnetic transition: in this case the small $\beta$ approximation fails to predict the simulated values. Larger values of $N$ and $T$ (not shown here) produce narrower error bars.

The general effect we see from Fig. \ref{fig:gaussauc_sim} is that higher variance of the $J$ and $h$ parameters leads to higher AUC values leaving all else unchanged (orange squares and yellow circles), while moving the means has little effect as long as the system is in its paramagnetic phase.

These results are easy to obtain thanks to the assumption that the model parameters $J$ and $h$ have Gaussian distributed entries, but in principle the distribution $\phi(g)$ can be derived also for other distributions, albeit probably requiring numerical solutions rather than the analytical ones we presented here.

\subsection{Further details on the DyEnKIM}

There are a couple of subtleties that need to be pointed out regarding the structure of the $B$ and $A$ parameters and of the Fisher Information $\mathcal{I}$ of the DyEnKIM, which are matrices rather than scalars as in the case of the DyNoKIM.

In order to make the estimation less computationally demanding in our example applications we  assume $A, B$ and $\mathcal{I}$ diagonal, disregarding the dependencies between time-varying parameters: this will likely make our estimates less precise, but it also reduces the number of static parameters to be inferred, letting us bypass model selection decisions which are outside the scope of this article.

As previously discussed there is also in this case the problem of identification for the averages of the components of $\boldsymbol{\beta}$, which we solve in the exact same way as we did for the DyNoKIM by dividing the values of each component by their sample mean while multiplying the associated static parameter by the same factor, again leaving the likelihood of the model unchanged, but setting a reference level for $\boldsymbol{\beta}$.

As a last remark, notice that the DyNoKIM and the DyEnKIM are equivalent when $h_0(t) = 0 \; \forall \, t$ and $\beta_{diag} = \beta_{off} = \beta_h = \beta$. For this reason we mainly present simulation results for the DyNoKIM alone to keep the manuscript concise, as we found no significant differences between the two models when it comes to the reliability of the estimation process.

\subsection{Consistency analysis for estimation}

We perform a consistency test on simulated data, aimed at understanding whether the two-step estimation procedure we outlined above is able to recover the values of the parameters of the model when the model itself generated the data.

\begin{figure}[h]
    \includegraphics[width=.5\linewidth]{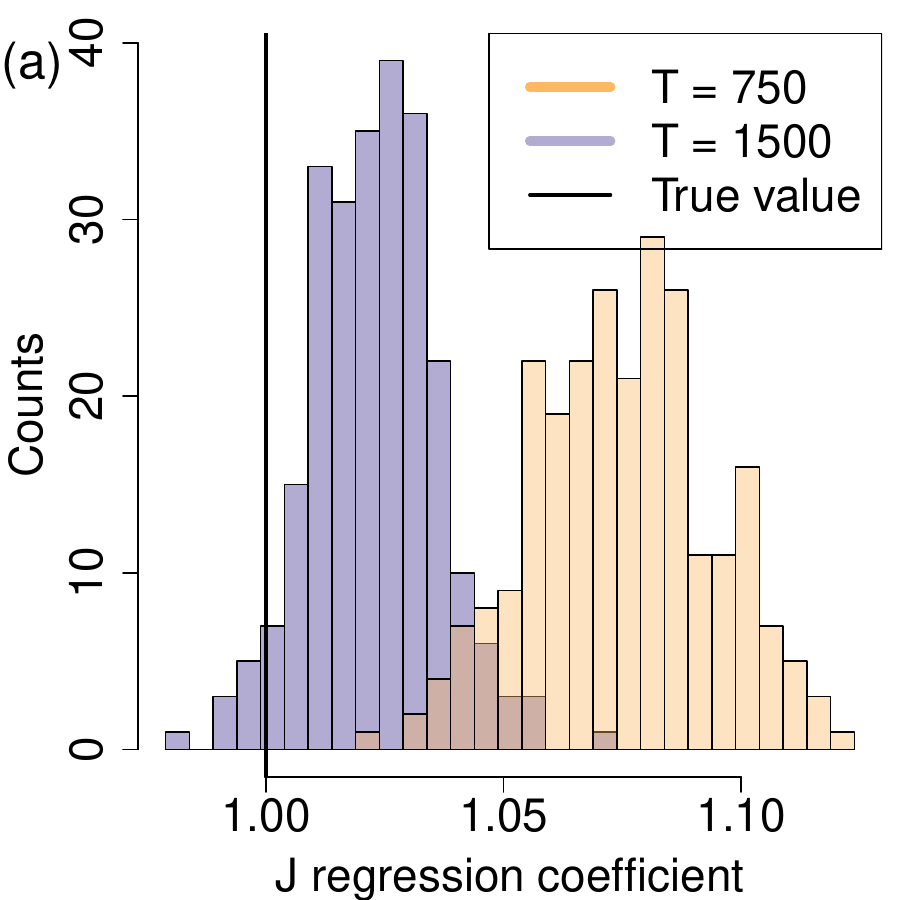}
    \includegraphics[width=.5\linewidth]{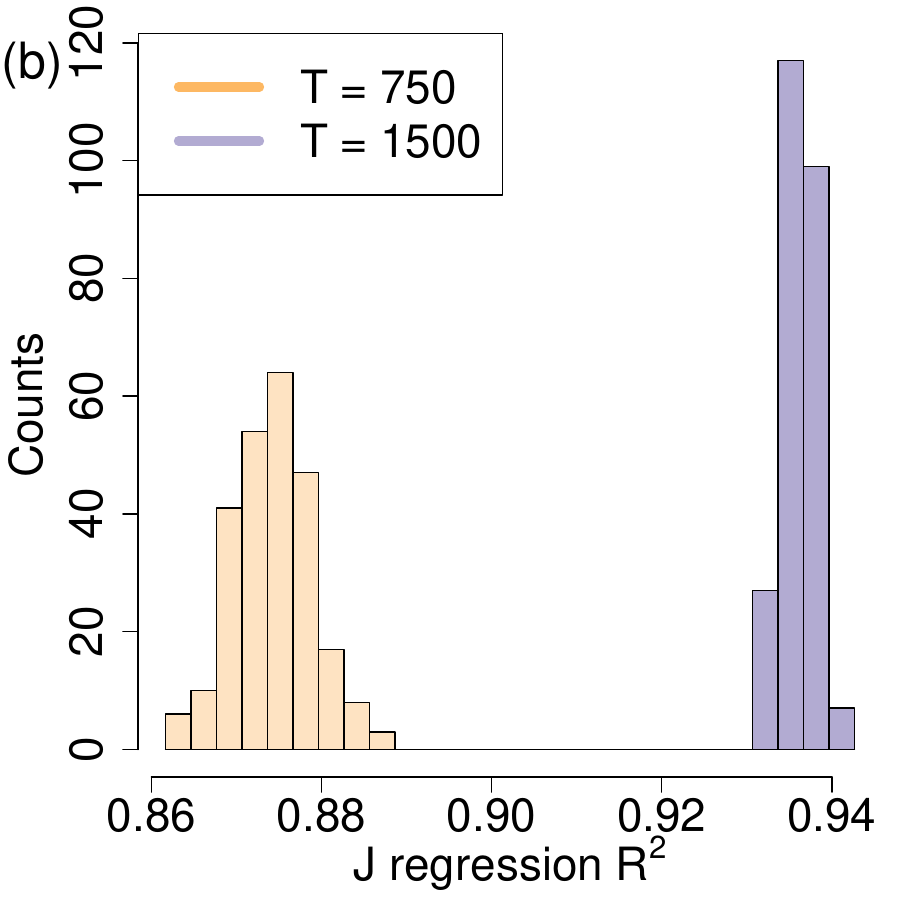}
    \caption{Consistency of the $J$ matrix estimation. (a) Histogram of linear regression coefficients $b$ between inferred and true values of $J_{ij}$ over 250 samples for $N = 50$, $T = 750$ and $T = 1500$; (b) Histogram of coefficients of determination ($R^2$) for the same set of models. The convergence of both values towards $1$ when increasing $T$ is a sign of consistency of the estimation.}
    \label{fig:consistency_J}
\end{figure}

Here we report results for simulations run with parameters $N=50$, $T=750$ or $T=1500$, $J_{ij} \sim \mathcal{N}(0, 1/\sqrt{N})$, $h_i = 0 \; \forall \, i$, $B = 0.95$ and $A = 0.01$. We see from Fig. \ref{fig:consistency_J} that the estimation of the elements of $J$ is indeed consistent: we estimate a linear regression model between the estimated and the true values of $J_{ij}$, namely $J^{est}_{ij} = a + b J^{true}_{ij} + \epsilon$, and plot the histogram of the values of $b$ and of the coefficient of determination $R^2$ of the resulting model from $250$ simulations and estimations ($a$ is consistently found to be very close to $0$ in all our simulations and for this reason we omit it). In the ideal case where for any $i,j$ $J^{est}_{ij} = J^{true}_{ij}$ one would have $b = R^2 = 1$, which is what we aim for in the limit $T \rightarrow \infty$. We see from our results that there is indeed a convergence of both values towards $1$ when increasing sample size, reducing both the bias and the variance of the regression parameters. 

\begin{figure}[h]
    \includegraphics[width=.5\linewidth]{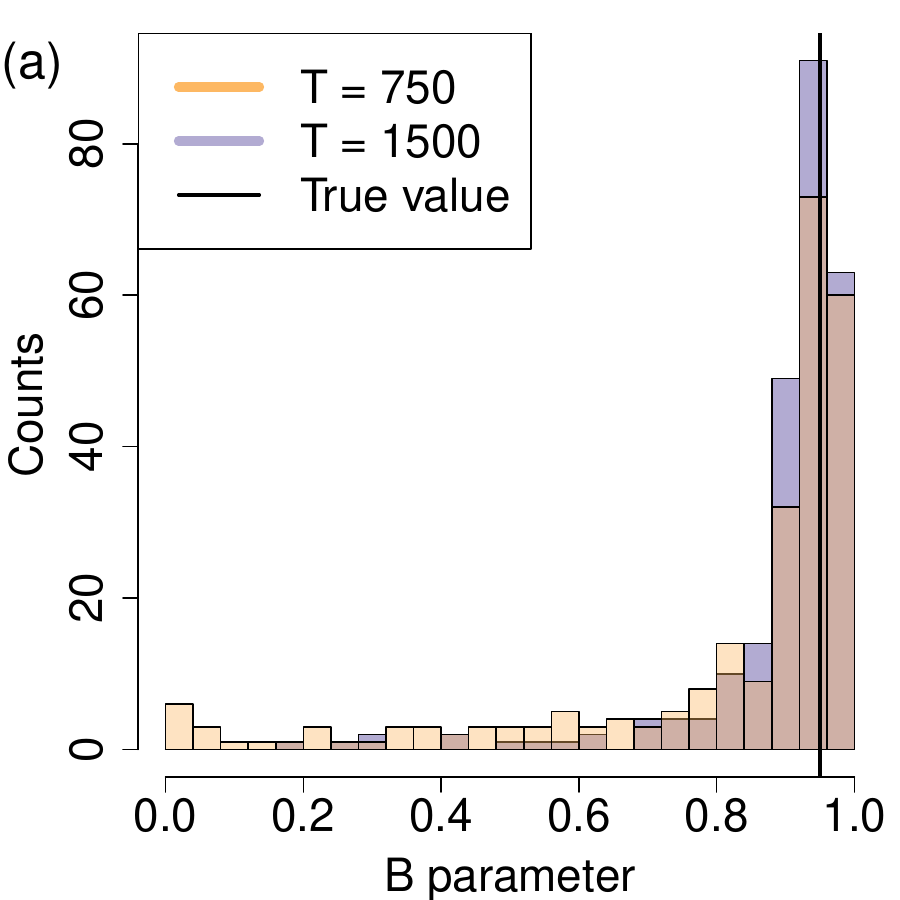}
    \includegraphics[width=.5\linewidth]{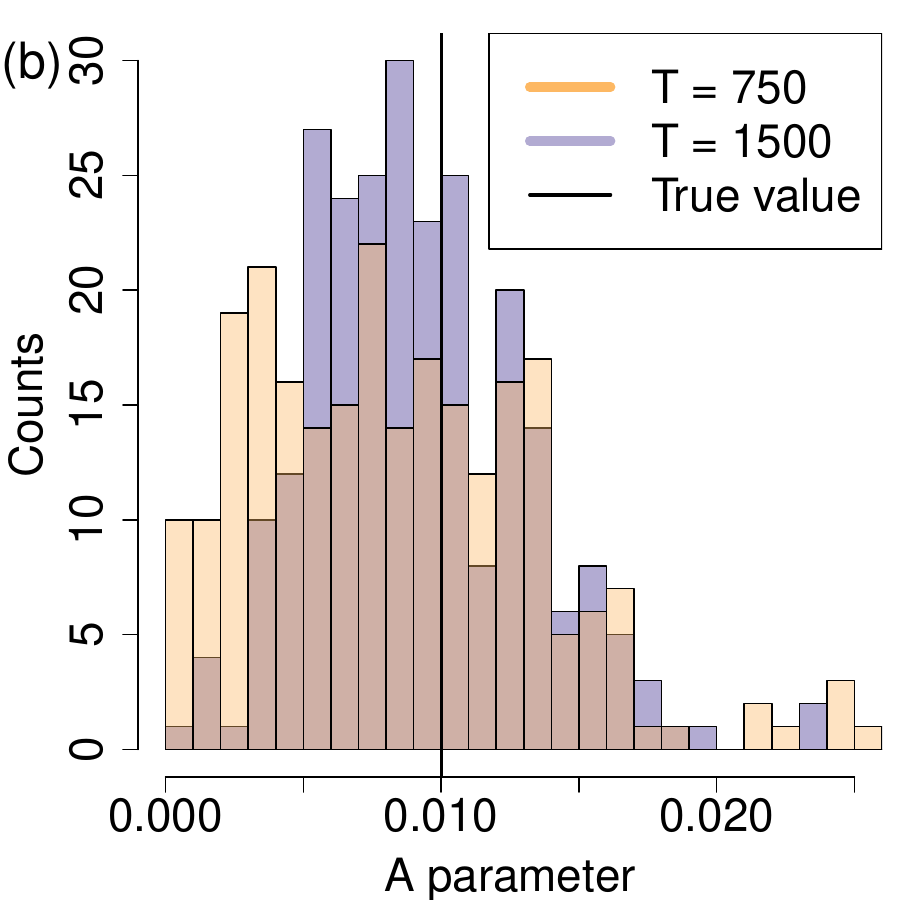}
    \caption{Consistency of the score-driven dynamics parameters. (a) Histogram of estimated values of $B$ over 250 samples for $N=50$, $T = 750$ and $T = 1500$; (b) Histogram of estimated values of $A$ over 250 samples for the same set of models. The convergence towards the true value by increasing $T$ is a sign of consistency of the estimation.}
    \label{fig:consistency_BA}
\end{figure}

Turning to the score-driven dynamics parameters $A$ and $B$, the situation does not change significantly. In Fig. \ref{fig:consistency_BA} we show the histograms of estimated values of $B$ and $A$ over $250$ simulations of $N=50$ variables for both $T = 750$ and $T = 1500$. It again appears clearly that when increasing the sample size the bias and variance of the estimators converge towards $0$, with the estimated parameter converging towards its simulated value. Thanks to these results we are able to confidently apply the two-step estimation method without the need to estimate all the parameters at once.

\begin{figure}[h]
    \centering
    \includegraphics[width=1\linewidth]{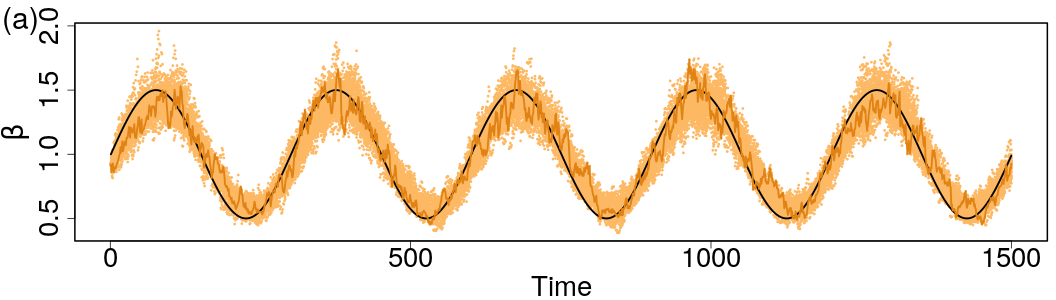}\\
    \includegraphics[width=1\linewidth]{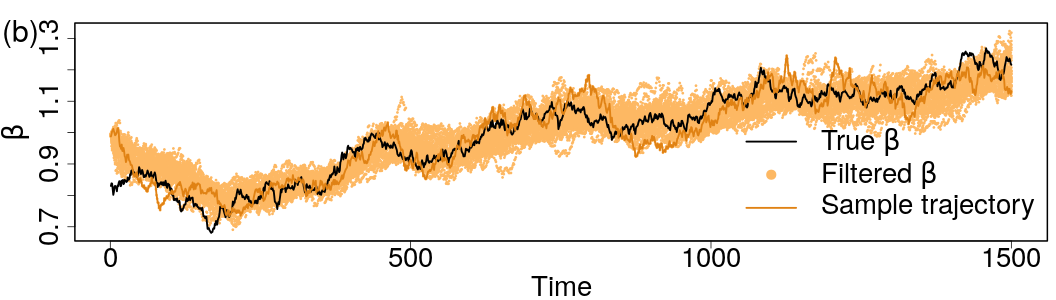}\\
    \caption{Simulation and estimation of a misspecified score-driven model over 30 simulations, with sample trajectories highlighted. (a) Deterministic $\beta$ following a sinusoidal function; (b) Stochastic $\beta(t)$ following an AutoRegressive model of order 1.}
    \label{fig:misspec_other}
\end{figure}

To add further evidence to what we presented in the main text, here we also report two additional figures regarding the filtering of misspecified $\beta(t)$ for the DyNoKIM and the DyEnKIM. In Fig. \ref{fig:misspec_other} we show two examples of misspecified $\beta(t)$ dynamics that are correctly recovered by the score-driven approach: the first is a deterministic sine wave function and the second is an AutoRegressive model of order $1$ (AR($1$)) which follows the equation

\begin{equation*}
    \beta^{AR}(t+1) = a_0 + a_1 \beta^{AR}(t) + \epsilon(t)
\end{equation*}
where $\epsilon (t) \sim \mathcal{N}(0, \Sigma^2)$ with parameters $a_0 = 0.005$, $a_1 = 0.995$, $\Sigma = 0.01$ so to have $\langle \beta^{AR} \rangle = 1$ and we select a simulation where $\beta(t) > 0$ $\forall$ $t$. In both cases we simulate 30 time series of length $T$ using the given values of $\beta(t)$ to generate the $s(t)$; given only the simulated $s(t)$ time series, the inference algorithm determines the optimal static parameters $A$, $B$ and $J$ and filters the optimal value of $\beta(t)$ at each time. We see that regardless of whether the underlying true dynamics is deterministic, stochastic, or more or less smooth the filter is rather accurate in retrieving the simulated values.

Regarding the DyEnKIM we want to show that different effects are correctly separated and identified when estimating the model on a misspecified data generating process. In fact while the consistency analysis largely resembles the one we reported for the DyNoKIM in Figures \ref{fig:consistency_J} and \ref{fig:consistency_BA} and for this reason we omit it, the effect of filtering multiple time-varying parameters is something that cannot be predicted by the simulations on the DyNoKIM alone.

\begin{figure}[h]
    \centering
    \includegraphics[width=1\linewidth]{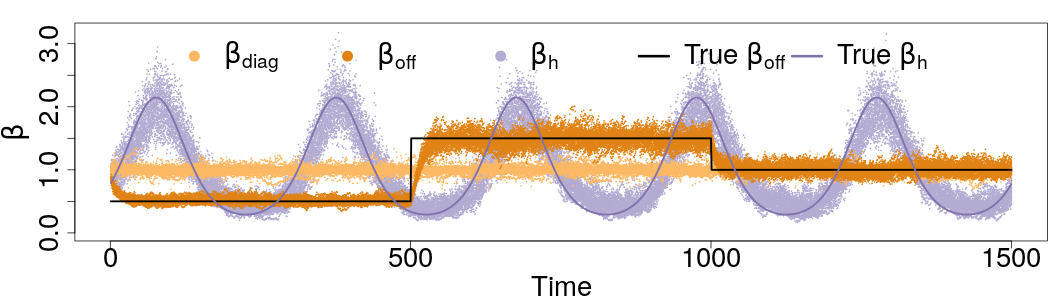}
    \caption{Estimation of $\beta_{diag}(t)$, $\beta_{off}(t)$ and $\beta_h(t)$ under model misspecification. The model was simulated with a constant $\beta_{diag}(t) = 1$, a piece-wise constant $\beta_{off}(t)$ and an exponentiated sinusoidal $\beta_h(t) = \frac{1}{\mathcal{J}_0(1)}\exp[\sin(\omega t)]$, with $\omega = 5 \frac{2 \pi}{T}$ and $\mathcal{J}_0$ the Bessel function of first kind of order 0 to normalize the mean. The points are the result of 30 different simulations and estimations, the lines show the values of $\beta_{off}$ and $\beta_h$ used to generate the data.}
    \label{fig:misspec_dye}
\end{figure}

In Figure \ref{fig:misspec_dye} we show the results when estimating the DyEnKIM on a dataset generated by a Kinetic Ising Model with time-varying $\beta_{diag}(t)$, $\beta_{off}(t)$ and $\beta_h(t)$ but where the dynamics of the parameters is predetermined instead of following the score-driven update rule. We arbitrarily choose to take a constant $\beta_{diag}(t) = 1$, a piecewise constant $\beta_{off}(t)$ and an exponentiated sinusoidal $\beta_h(t) = \exp[\sin (\omega t)]$, with $\omega = 5 \frac{2 \pi}{T}$, $T=1500$ and $N=30$. The results show that the filter works correctly and that the different time-varying parameters are consistently estimated, regardless of the kind of dynamics given to each of them.

\newpage

\newpage

\newpage

\newpage

\newpage

\newpage

\newpage 

\newpage

\section{Additional results on applications}

\subsection{Additional results on neural population data}

In this section we provide further results on the application of the KIM to the neural population data. In particular we discuss the possibility to use the DyEnKIM to test the significance of the elements of $J$ fitted over multiple experiments and expand on the comparison between the static parameters KIM and the score-driven version.

\begin{figure}[h]
    \centering
    \includegraphics[width=1\linewidth]{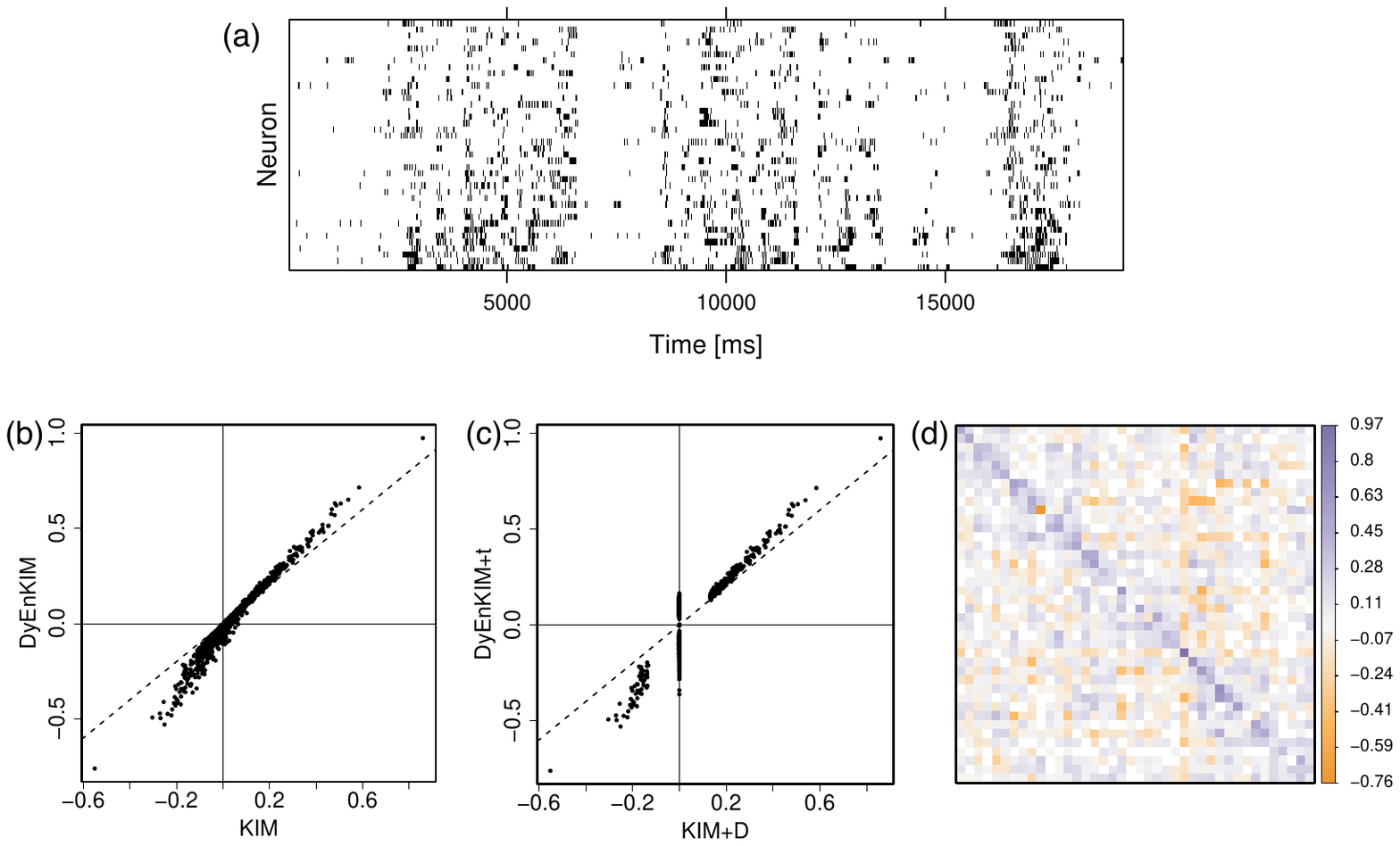}
    \caption{Additional results on the neuron spike data. (a) Raster plot for one sample in the data. Time is on the x-axis, neurons on the y-axis, a black dot at (t,i) indicates a spike from neuron $i$ at time $t$; (b) Comparison between fitted values of $J_{ij}$ using a KIM over the complete dataset (x-axis) and the average fitted values of $J_{ij}$ with DyEnKIM correction, $\overline{J}_{ij} = \frac{1}{M} \sum_i^M J_{ij}^{(k)}$, where $J_{ij}^{(k)}$ is the $J$ matrix fitted on sample $k$ and $M$ is the total number of experiments (y-axis). A dashed line is traced on the diagonal as guide to the eye; (c) The same as panel b after the Decimation pruning technique has been applied to the KIM and the t-test pruning has been applied to the DyEnKIM; (d) Visualization of the DyEnKIM $\overline{J}$ after the t-test pruning has been applied. White squares correspond to non-validated interactions.}
    \label{fig:neurons_extra}
\end{figure}

In Figure \ref{fig:neurons_extra}a we show an example of the analyzed data in the form of a raster plot of one of the experiments. It appears clear that the dynamics of spikes is bursty and there are non-negligible auto- and cross-correlation effects among neurons, likely driven by the external stimulus of the video \cite{tkavcik2014searching}. 

As mentioned in the section on the DyNoKIM in the main text, it is possible to use the filtered time-varying parameters to correct the values of $J$ from misspecification error. The same can be done with the DyEnKIM, correcting the elements of $J$ by a factor given by the sample mean of $\beta_{diag}$ and $\beta_{off}$ and the external fields $h$ by the sample mean of $\beta_h$. In Figure \ref{fig:neurons_extra}b we show a scatterplot of the values of $J_{ij}$, comparing the parameters fitted with a KIM on all available data (x-axis) with the average value $\overline{J}_{ij} = \frac{1}{M} \sum_k^M J_{ij}^{(k)}$, where $J_{ij}^{(k)}$ is the value fitted with the DyEnKIM correction on experiment $k$. It is clear that, as in the case of DyNoKIM shown in Figure 2 of the main text, the correction - which in this case only affects off-diagonal elements as $\beta_{diag} = 1$ - tends to increase the absolute value of $J_{ij}$ with respect to the static KIM version.

Pruning irrelevant parameters is central to the definition of meaningful statistical models. Here we propose two alternative methods, Decimation and t-testing; the first is standard in the literature on Kinetic Ising Models \cite{decelle2015inference}, whereas the second exploits the repeated experiments in the data to compare parameters fitted on different samples and assess their significance by means of a t-test. For Decimation we refer the interested readers to the original paper introducing it \cite{decelle2015inference}. As an alternative in cases where multiple repetitions of the experiment are available, as is the case for our neuron spike dataset, it is possible to fit a DyEnKIM for each of the $M$ experiments and then use a Student's t-test on the set of values $J_{ij}^k$, $k=1, \dots, M$ to test whether their average is significantly different from 0.
In Figure \ref{fig:neurons_extra}c we show a comparison between these two methods, with Decimation applied to the KIM $J$ and the t-test used to validate the DyEnKIM result, adopting a Bonferroni correction for multiple hypothesis testing at the $p < 0.01$ level. In our case, the Decimation approach selects less elements of $J_{ij}$ as significant, whereas the t-test appears to be less specific or more sensitive. This difference is possibly related to Decimation being a likelihood-based method, which may suffer from misspecification in case the data generating process is not a KIM, but answering this question goes beyond the scope of this paper. Finally, in Figure \ref{fig:neurons_extra}d we report a visualization of the t-tested DyEnKIM $\overline{J}$ matrix. The diagonal elements are largely positive, indicating significant autocorrelation in spiking dynamics (as would be expected by a visual inspection of the raster plot), whereas off-diagonal elements are generally smaller in absolute value and both positive and negative, albeit significantly different from 0.

\subsection{Results on the application of DyEnKIM on regular trading days}

\begin{figure}[h]
    \includegraphics[width=.5\linewidth]{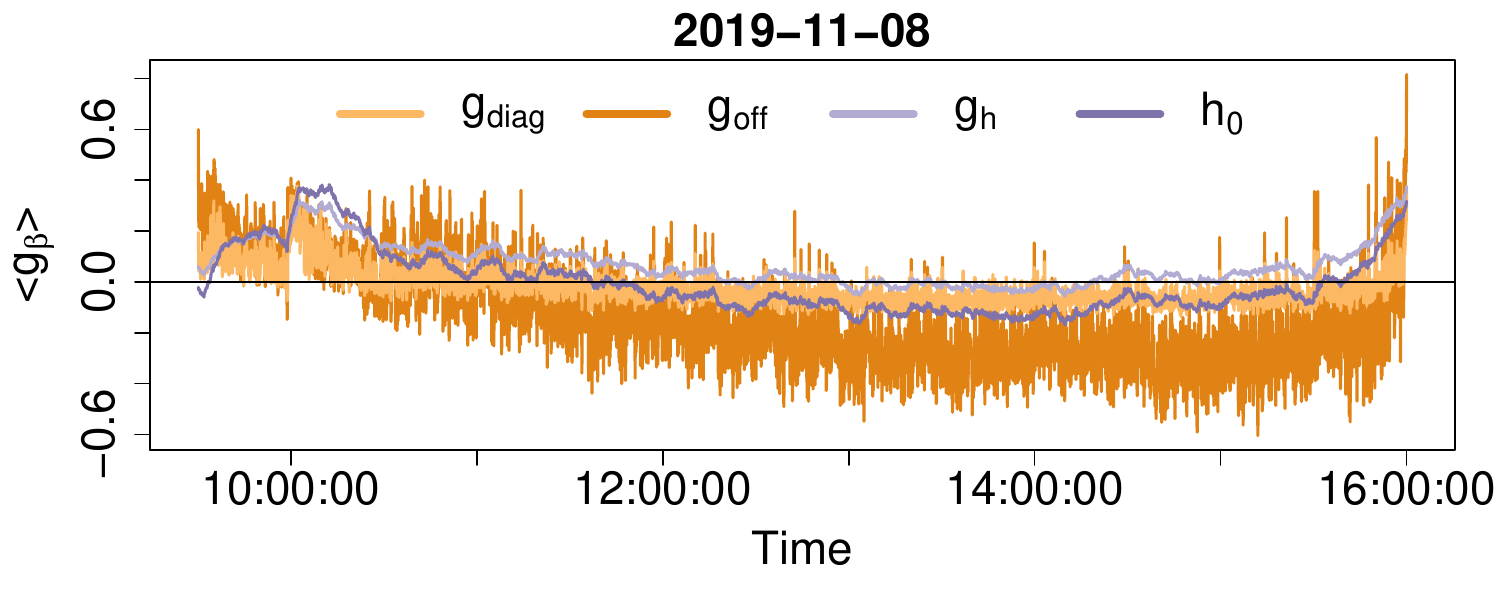}
    \includegraphics[width=.5\linewidth]{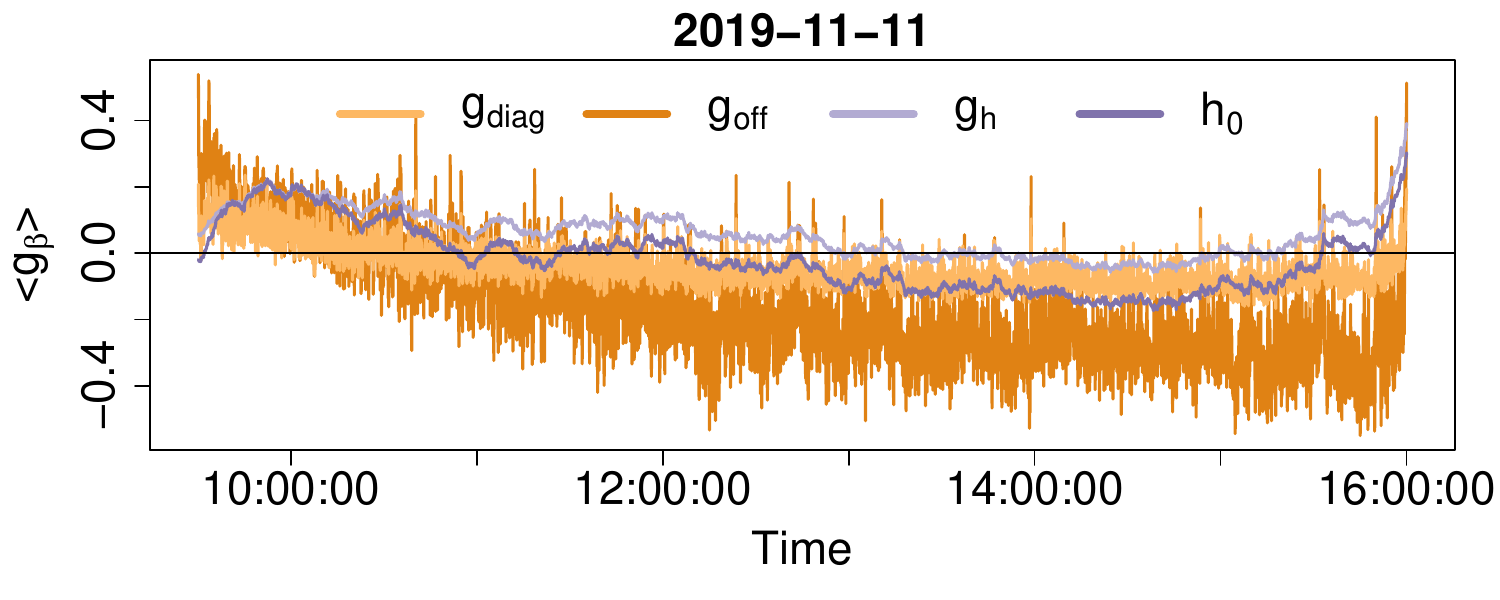}
    \includegraphics[width=.5\linewidth]{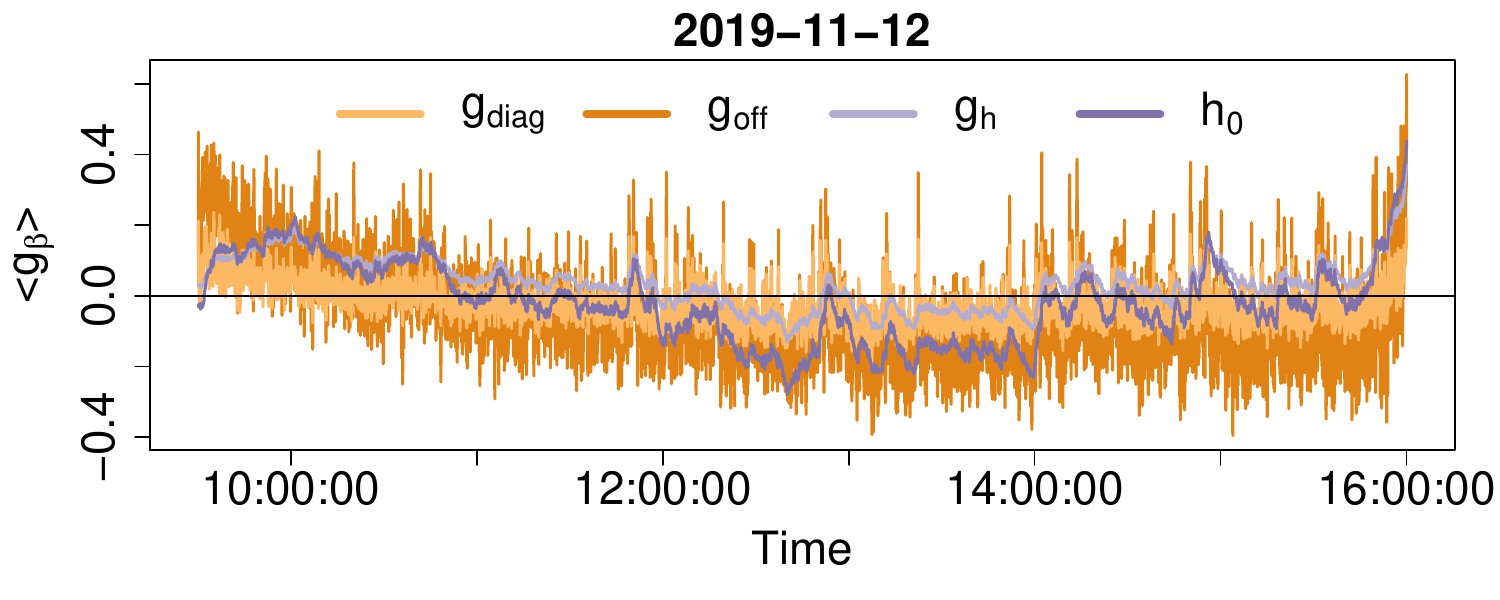}
    \includegraphics[width=.5\linewidth]{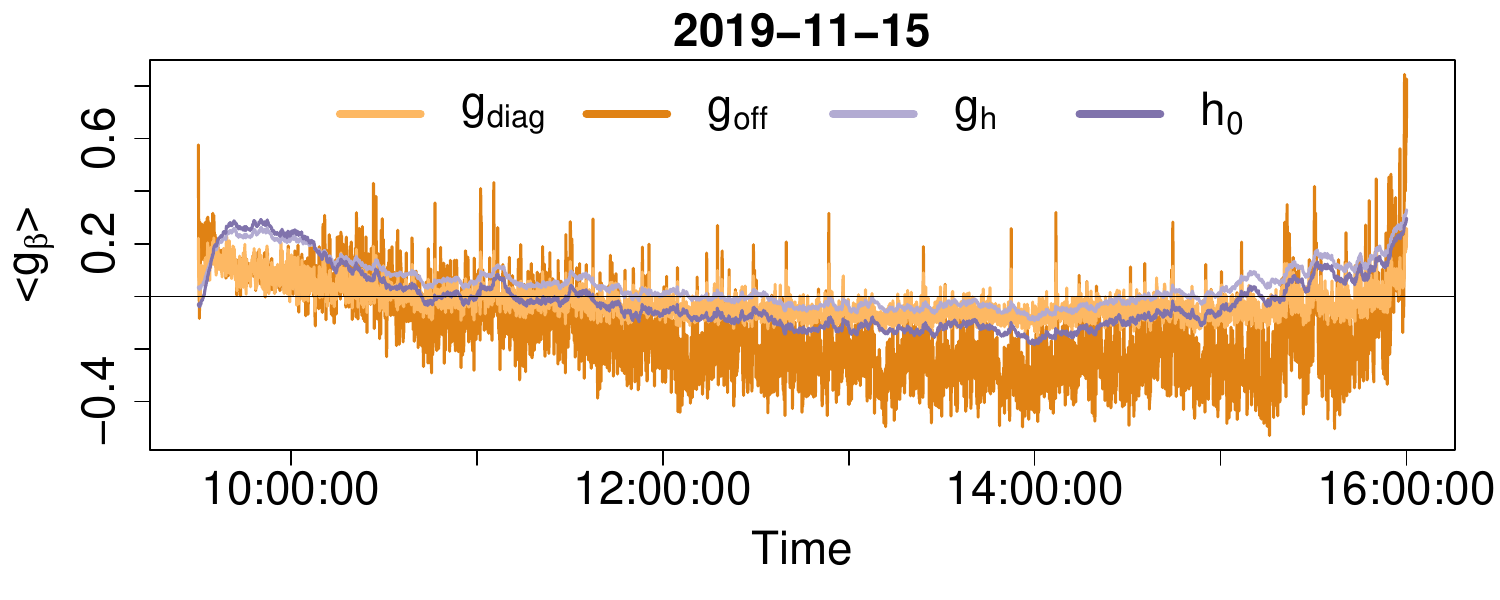}
    \includegraphics[width=.5\linewidth]{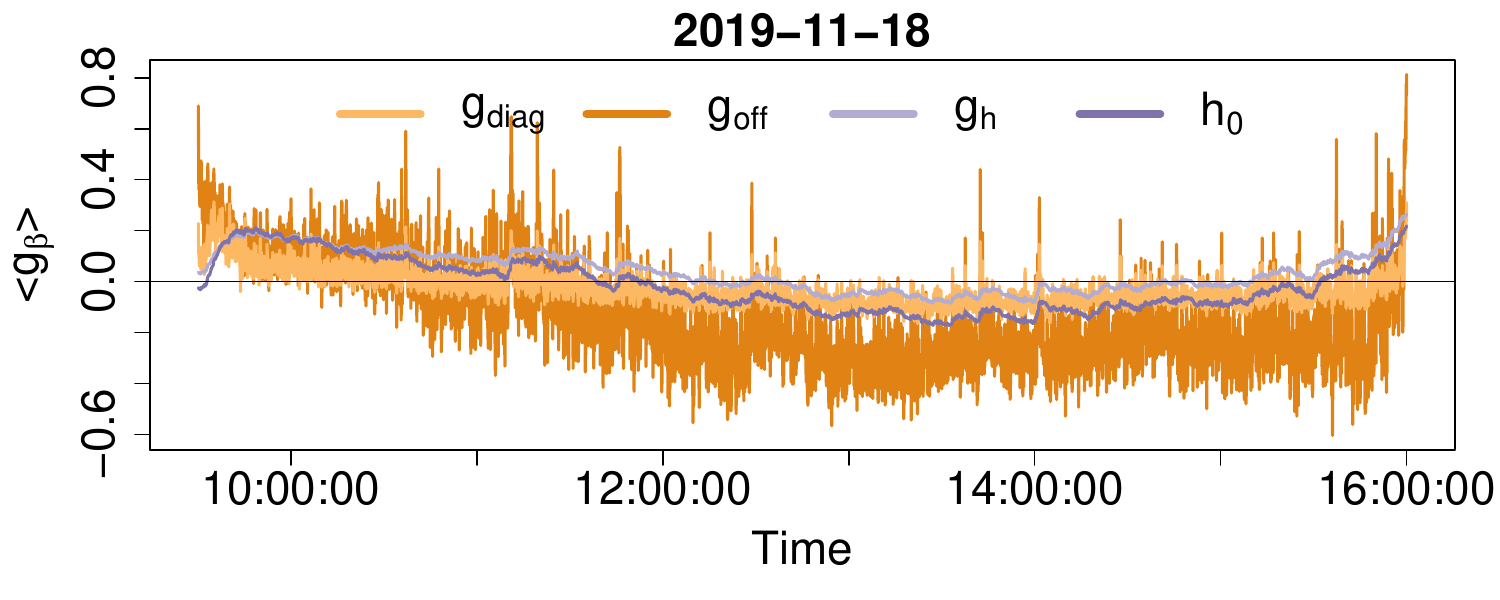}
    \includegraphics[width=.5\linewidth]{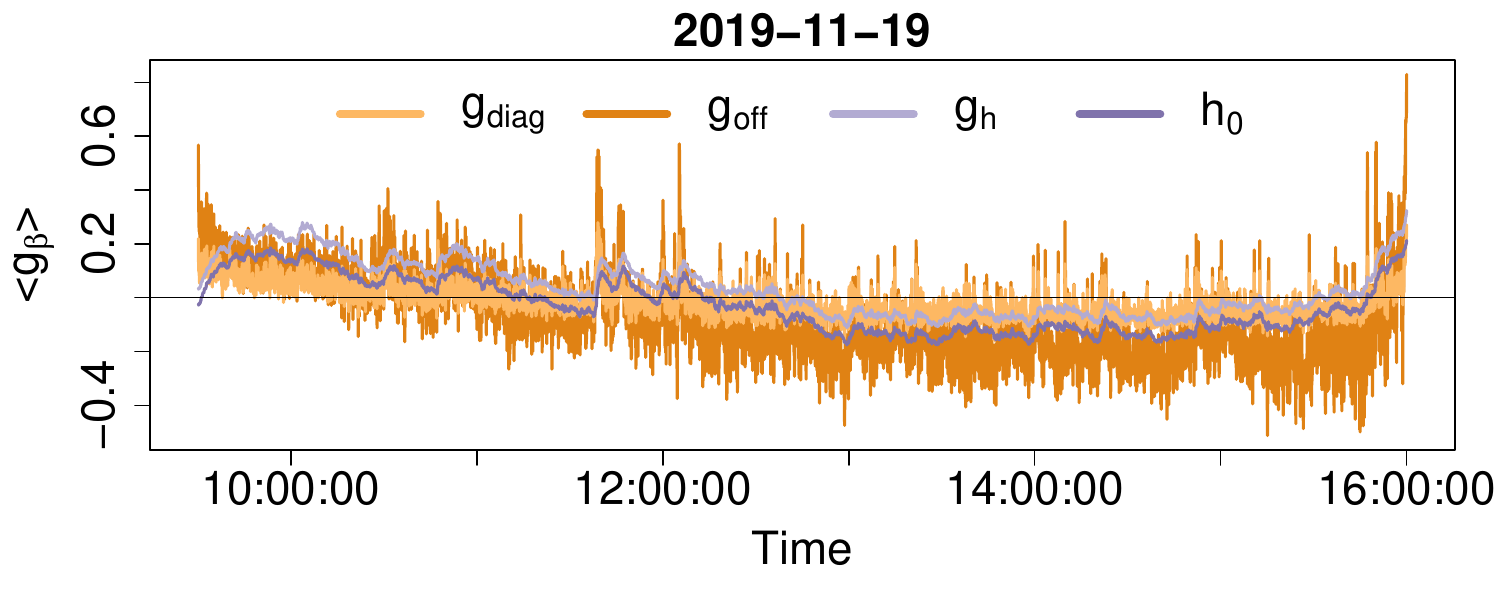}
    \caption{Values of 
    $\langle g_{diag} \rangle (t)$, $\langle g_{off} \rangle (t)$, $\langle g_{h} \rangle (t)$ and $h_0(t)$ during six days in November 2019, when no abnormal event was recorded. The usual U-shaped pattern of intraday volatility and volume is observed.}
    \label{fig:regular_tradingday}
\end{figure}

As a comparison with the observations reported in the main text regarding particular events affecting stock markets, such as the Flash Crash and the FOMC announcement of July 31st 2019, here we briefly discuss observations for a regular trading day where nothing as exceptional happened. In Figure \ref{fig:regular_tradingday} we replicate the plot shown in Figure 5 of the main text for six days in November 2019. Here we see that the $J$-related components of effective fields $\langle g_{diag} \rangle(t)$ and $\langle g_{off} \rangle(t)$ show a U-shaped pattern throughout the trading day, having higher values at the opening and closing, while the $h$-related $\langle g_h \rangle $ only shows an increase towards the end of the day. The $h_0$ parameter, which captures the average exogenous price activity across all stocks, shows itself a U-shaped pattern which is more pronounced at closing, consistent with the intraday pattern typical of traded volume. The consistency of this result throughout these relatively uneventful days thus reinforces the qualitative description provided by the DyEnKIM for the turbulent events analyzed in the main text.

\subsection{Details on the Relation Between KIM and TERGM}
Here we discuss the relation between the Kinetic Ising Model (KIM) and the Temporal Exponential Random Graph (TERGM) of \cite{Hanneke_tergm_10.1214/09-EJS548}, and show how our score driven extensions can be seen as time varying parameters extensions of TERGM. 

Considering only lag 1 dependencies, a TERGM is defined by the following probability mass function 
\begin{equation}\label{eq:TERGM_def}
    P(G(t)|G(t-1), \theta ) = \frac{e^{\sum \theta_l q_l(G(t),G(t-1))}}{K(\theta)},
\end{equation}

where the functions $q_l(G(t),G(t-1))$, called network statistics, are defined to investigate the determinants of the network's dynamics, and $K(\theta)$ is a normalization coefficient, known as partition function. Examples of network's statistics are 
$q_{stab}(G(t), G(t-1)) = \sum_{ij} G_{ij}(t) G_{ij}(t) + (1 - G_{ij}(t))(1- G_{ij}(t-1)) $, 
that captures links' stability, and
$ q_{dens}(G(t)) = \sum_{ij} G_{ij}(t)$, 
related to network's density.

As mentioned in the main text, we can easily map each entry of the adjacency matrix into a spin and associate a sequence of spins to a discrete time temporal network. Indicating by $vec(G)$ the vectorized version of matrix $G$,  the mapping can be summarized by the relation 
$$
vec(G(t))_{k} =  1/2 + s_{k}(t)/2
$$ and the KIM is equivalent to the following version of the TERGM $$
P(s(t)|s(t-1), \theta ) = \frac{e^{ \sum_{ij} J_{ij} s_{i}(t-1) s_{j}(t)  +  \sum_i h_{i} s_i(t)  } }{Z(J, h)} = \frac{e^{ \sum_{ab} \theta^{(1)}_{ab} vec(G(t-1))_{a} vec(G(t))_{b}   +  \sum_{a} \theta^{(2)}_a vec(G(t))_{a} - \sum_{a} \theta^{(3)}_{a} vec(G(t-1))_{a}} }{K(\theta)},
$$
where we omitted the constant terms, not depending on the adjacency matrix, that have been absorbed in the normalization constant.
Hence, KIM is equivalent to a TERGM having three kinds of network statistics: first, the set of all possible lagged interactions $q^{(1)}_{ab} =  vec(G(t-1))_{a} vec(G(t))_{b}$ between pairs of links, each appearing in this specification of Eq. \ref{eq:TERGM_def} with a parameter  $\theta^{(1)}_{ab} = 4 J_{ab}$; second, a term associated to the probability of each link to be observed, $q^{(2)}_{a} = vec(G(t))_{a}$, with  $\theta^{(2)}_{a} = 2 ( h_a-1 ) $; and the last group of statistics, $q^{(3)}_a = vec(G(t-1))_{a}$, is related with the presence or absence of each link at the previous time step with parameters $q^{(3)}_a = 2\sum_b J_{ab}$.  

The DyNoKIM allows us to consider time varying $\theta_i$ and to estimate the forecast accuracy of the model at each time step, as showed with the link prediction example presented in the main text. 

Interestingly,  a wide range of TERGM specifications can be mapped to the KIM. As a simple example, let us consider a TERGM with two terms only 
\begin{equation} \label{eq:example_TERGM}
    P(G(t)|G(t-1), \theta )  = \frac{e^{ \sum_{ij} \left[ \theta_{dens} G_{ij}(t) +  \theta_{stab} ( G_{ij}(t) G_{ij}(t-1) + (1 - G_{ij}(t))(1- G_{ij}(t-1)) ) \right] } }{K(\theta)}.
\end{equation}
This can be rewritten as
$$
P(s(t)|s(t-1), \theta ) = \frac{e^{ N \theta_{dens}/2 +  \sum_{k} \left(s_{k}(t) \theta_{dens}/2  +  s_{k}(t) s_k(t-1) \theta_{stab}/2  \right)  } }{K(\theta)},
$$
which is exactly equivalent to a KIM
restricted to have just two parameters $J_{ij} = J_{diag} = \theta_{stab}/2 \; \forall i, j$ and $h_i = h_0 = \theta_{dens}/2  \; \forall i$, and, absorbing the constants in the normalization function, we have
$$
P(s(t)|s(t-1), J_{diag}, h_0 ) = \frac{e^{ \sum_{k} s_{k}(t) h_0  +  s_{k}(t) s_k(t) J_{diag}  } }{Z(\theta)}.
$$
If we consider the DyNoKIM extension of such a restricted KIM we obtain 
\begin{equation} \label{eq:restr_DyNoKIM}
   P(s(t)|s(t-1), J_{diag}, h_0, \beta(t)) = \frac{e^{ \beta(t) \sum_{k}\left( s_{k}(t) h_0  +  s_{k}(t) s_k(t-1) J_{diag}  \right) } }{Z(\theta)}
\end{equation}
that is effectively an extended version of the initial TERGM. 
Moreover, it is easy to see that the DyEnKIM results in the following 
\begin{equation} \label{eq:restr_DyEnKIM}
     P(s(t)|s(t-1), J_{diag}, h_0, \beta(t)) = \frac{e^{ \sum_{k}\left( s_{k}(t) \beta_{h}(t) h_0  +  s_{k}(t) s_k(t-1) \beta_{diag}(t) J_{diag}  \right) } }{Z(\theta)}.
\end{equation}
that maps to a version of \eqref{eq:example_TERGM} with dynamical parameters $\theta_{dens}(t)$ and $\theta_{stab}(t)$ evolving independently. We believe this observation is very relevant as it is an extension of the TERGM at hand to its version where each parameter is allowed to follow its own evolution, potentially unrelated to the others. This is a different evolution from the DyNoKIM's one, as the latter is driven by a single $\beta(t)$ and maps into comoving TERGM parameters. Indeed, also in this context, the two models have different purposes and different applicability. TERGM extensions resulting from DyNoKIM allow us to quantify forecast accuracy, similarly to what showed in the main text, while DyEnKIM, similarly to what we discussed in the applications presented and in numerical simulations, allows for a decoupling of the temporal relevance of different network statistics.

As a final remark, we point out that the class of TERGMs that can be mapped into KIMs, and benefit of the corresponding score driven extensions, is not restricted to cases with linear dependency on the lagged adjacency matrix. In fact, we can also consider network's statistics depending on products of lagged matrix elements, e.g. $h(G(t),G(t-1)) = \sum_{ijk} G_{ik}(t)G_{ij}(t-1)G_{ik}(t-1) $, as long as they depend linearly on $G(t)$.  A TERGM with such statistics can be mapped in a KIM with the addition of predetermined regressors and is easily extended, for example, to the corresponding DyNoKIM version. For example, all the statistics discussed as explicit examples in \cite{Hanneke_tergm_10.1214/09-EJS548} take this form, and can be mapped into a KIM.  Although a full characterization of the set of TERGM's specifications that can be mapped into a KIM lies outside the scope of this work, we suspect it to be very large, and potentially include all statistics commonly used in practice. 

In summary, as we have shown that KIM belongs to the TERGM family, and its score driven extensions result in extensions of the corresponding TERGM. That is the case for both DyNoKIM and DyEnKIM, as we showed explicitly for a simple TERGM specification. We believe that our findings open up a vast space of potential applications of score driven KIM to temporal networks, and raise interesting theoretical questions on the possibility of mapping a generic TERGM into a KIM, that we leave for future explorations.

\newpage

\newpage

\end{appendices}

\end{document}